\documentclass[12pt]{article}

\usepackage{pdfpages}
\usepackage{hyperref}
\usepackage{graphicx}
\usepackage{dcolumn}
\usepackage{bm}
\usepackage{verbatim} 
\usepackage{lscape} 
\usepackage{amsmath}
\usepackage{amssymb}
\usepackage{appendix}
\usepackage{multirow}
\usepackage[margin=1.0in, top=1.0in, bottom=1.0in]{geometry}

\begin{document}
\large


\title{
Reconstruction: Rational Approximation of the Complex Error Function 
and the Electric Field of a Two-Dimensional Gaussian Charge Distribution
}
\author{
\centerline{John Talman$^{1}$, Yuko Okamoto$^{2}$, and Richard Talman$^{3}$} \\ \\
\centerline{$^{1}$UAL Consultants, 327 Savage Farm Drive, Ithaca, N.Y. 14853, USA} \\
\centerline{$^{2}$Information Technology Center and Global Engagement Center, } \\
\centerline{Nagoya University, Furo-cho, Chikusa-ku, Nagoya, Aichi 464-8601, Japan} \\
\centerline{$^{3}$Laboratory for Elementary-Particle Physics, } \\
\centerline{Cornell University, Ithaca, N.Y. 14853, USA}
}

\maketitle

\begin{abstract}
This paper resurrects and archives an unpublished original Cornell
Laboratory of Nuclear Studies report by Yuko Okamoto and Richard
Talman, “Rational Approximation of the Complex Error Function and
the Electric Field of a Two-Dimensional Gaussian Charge Distribution”
CBN 80-13, dating from September 1980, during the start-up period
of the CESR-CLEO e$^+$e$^-$ collider. This code has played a significant
role in the calculation of the beam-beam interaction (in particular the
beam-beam tune shift) for subsequent storage ring colliders. Electronic
access to the (refactored) original codes is provided by active links.
\end{abstract}

     
\tableofcontents

\section{Introduction\label{sec:Introduction}}
Various individuals have suggested we re-create and archive an unpublished 1980 
Cornell Laboratory of Nuclear Physics 
``CBN 80-13'' report, with the title given above, written at that time by two of 
the present authors (Y.O. and R.T.), 
in order to make it more readily and securely accessible and useable than at present.  The main body of the present
article is a faithful, page-by-page (latex-produced) reproduction of the original report.  This includes
all tables and includes hand-written annotations (by Y.O.) comparing the precision obtained, as matched with 
earlier sources (referenced in the report), especially in ``difficult'' regions of parameter space.  

The original report is copied verbatim in Section~\ref{sec:OriginalReport}.  
Though ideal for retention of chronology, this text is
not at all convenient for modern day application of the original (Fortran) codes.  For this reason the original
codes have been refactored and made available online, using the active links given in Section~\ref{sec:ActiveLinks}.
Because of the inherent backward compatibility of Fortran it was possible to do this without much risk of
introducing errors in this process.  To confirm faithful reproduction, and to provide a benchmark for subsequent
reproductions, one of the original tables, Table 1, WEXCT, has been reproduced in Section~\ref{sec:WEXCT}. 

\section{Active Links\label{sec:ActiveLinks}}
For more information about this report and for accessing the original complex error 
function Fortran code, the online ``Okamoto'' repository can be accessed from a web browser
at the github URL\\
 ``https://github.com/jtalman/ual1/tree/master/Okamoto/fortran''\\
which contains refactored Fortran code and other related material.

\emph{The following ``Fortran code'' active link to this ``git'' repository may have been deactivated by the 
arXiv or by your mail system.  If so, it is necessary to access ``Fortran code'' at the github URL given above. 
The same is true for the following ``Accelerator Simulation Course'' link.}

\href{https://github.com/jtalman/ual1/tree/master/Okamoto/fortran}{Fortran code}

\bigskip
\noindent
The following active link points to course notes, in the same github repository, for a UAL (Unified Accelerator 
Libraries) Simulation course at the 2005 U.S. Particle Accelerator Course given at Cornell by Nikolay Malitsky and 
Richard Talman.  Chapter 9, ``Colliding Beams''  starting on page 145, explains why the complex error function is
needed for simulating the beam-beam interaction in colliding beam storage rings. Section 9.6 and 9.7 provide some
of the evolution of the Pade code that had been developed since its original application during the commissioning 
of the CESR-CLEO e$^+$e$^-$ Collider beginning in 1980, when the code being replicated in the present paper was 
generated.

\href{https://github.com/jtalman/ual1/blob/master/examples/ETEAPOT/legacyAndFoundation/CornellUSPAS.pdf}{Accelerator Simulation Course}

\clearpage
\section{Original report\label{sec:OriginalReport}}

\begin{abstract}
To simulate the beam-beam interaction one needs efficient formulae for the 
evaluation of the electric field of a two-dimensional Gaussian charge
distribution
which can be expressed in terms of the complex error function $w(z)$.  This 
paper shows how to approximate $w(z)$ by a set of rational functions.  The 
percent error of the approximation is extremely small ($\sim 10^{-4}$\% except 
near the real axis). Computer programs to evaluate $w(z)$ and the electric field 
are also provided.
\end{abstract}

\subsection{Introduction}
For the simulation of the beam-beam interaction one needs to evaluate the 
electric field of a two-dimensional Gaussian charge distribution. The electric 
field at the position $(x,y)$ has been found by M. Bassetti and 
G.A. Erskine \cite{Bassetti} to have the following form: \cite{Bassetti-typo}
\begin{align} 
E_x &= \frac{Q}{2\epsilon_0\sqrt{2\pi(s_x^2-s_y^2)}}\,\Im\,
\Bigg(
w\bigg(\frac{x+iy}{\sqrt{2(s_x^2-s_y^2)}}\bigg)
 - 
e^{-\big(\frac{x^2}{2s_x^2} + \frac{y^2}{2s_y^2}\big)}\,w\bigg(\frac{x\frac{s_y}{s_x} + 
iy\frac{s_x}{s_y}}{\sqrt{2(s_x^2-s_y^2)}}\bigg) \Bigg),   
\label{eq:E_x} \\
E_y &= \frac{Q}{2\epsilon_0\sqrt{2\pi(s_x^2-s_y^2)}}\,\Re\,
\Bigg(
w\bigg(\frac{x+iy}{\sqrt{2(s_x^2-s_y^2)}}\bigg)
 - 
e^{-\big(\frac{x^2}{2s_x^2} + \frac{y^2}{2s_y^2}\big)}\,w\bigg(\frac{x\frac{s_y}{s_x} + 
iy\frac{s_x}{s_y}}{\sqrt{2(s_x^2-s_y^2)}}\bigg) \Bigg),  
\label{eq:E_y} 
\end{align}
where $\Im$ and $\Re$ stand for imaginary part and real part, respectively, 
$Q$ is a constant with a dimension of electric charge, $\epsilon_0$
is the electric permittivity of free space, $s_x$ and $s_y$ ($s_x>s_y$ assumed) 
are the standard deviations of the charge distribution in the $x$ and $y$ directions, 
respectively, and $w(z)$ is the complex error function \cite{Abramowitz} defined by 
\begin{equation}
w(z) = e^{-z^2}\,
\Big( 1 + \frac{2i}{\sqrt{\pi}}\,\int_0^z\,e^{u^2}\,du\Big).
\label{eq:w_z}
\end{equation}

We shall approximate $w(z)$ by rational functions so that a computer can 
\emph{quickly} handle the evaluation of $w(z)$ and thus the electric field 
of a two-dimensional Gaussian charge distribution.  Though we were originally
interested in an approximation good to within 1\% error, the result turned 
out to be a much better approximation.  We note that after the approximation 
of $w(z)$ the only transcendental function in (\ref{eq:E_x}) and  (\ref{eq:E_y}) 
which spend a  longer computing time than rational functions are the 
exponential factors. We also note that by the symmetry properties 
of $w(z)$ \cite{Abramowitz} it suffices to approximate $w(z)$ only in the first 
quadrant of the complex plane.

\subsection{Pad\'e Approximation}
We shall briefly describe how the Pad\'e approximation is done first, then
apply the approximation to the function $w(z)$. 

Suppose that we have a complex-valued function $f(z)$ which is analytic at a 
point $z_0$ and suppose that we want to approximate it around $z_0$ by a rational
function of the form 
\begin{equation}
f_{\rm Pade}(z) = \frac{\displaystyle{\sum_{k=0}^M}\,c_k(z-z_0)^k}{1 + 
\displaystyle{\sum_{k=1}^N}\,d_k(z-z_0)^k} \, , 
\label{eq:Pade.1}
\end{equation}
where $c_k, d_k \in \mathfrak{C}$ are unknown (possibly complex) coefficients 
to be determined.  Note: We must have $d_0\neq0$ because $f(z)$ is well-behaved
at $z_0$. We may set $d_0=1$. For, otherwise, we can always divide both the 
numerator and denominator by $d_0$.

Here we choose $M$ and $N$ according to how much accuracy we need. In order to 
determine the coefficients $c_k$ and $d_k$ we impose a condition on $f_{\rm Pade}$:
\begin{equation}
f - f_{\rm Pade} = A_1(z-z_0)^{M+N+1} + A_2(z-z_0)^{M+N+2} + \cdots,
\label{eq:Pade.2}
\end{equation}
where  $A_1, A_2, \cdots \in\mathfrak{C}$ are some constants. That is, the error
introduced by the approximation at $z$ with $|z-z_0|<1$ is of the order of 
$|z-z_0|^{M+N+1}$ and very small if $M$ and $N$ are large. Since $f$ is analytic 
at $z_0$, we have a Taylor series at $z_0$:
\begin{equation}
f(z) = \sum_{j=0}^{\infty}\,a_j(z-z_0)^j;\quad a_j\in\mathfrak{C}.
\label{eq:Pade.3}
\end{equation}
Then using (\ref{eq:Pade.3}) for $f$ in (\ref{eq:Pade.2}), multiplying 
both sides of (\ref{eq:Pade.2}) by the denominator of $f_{\rm Pade}$, and equating 
the coefficients of the powers of $(z-z_0)$ in both sides of the equation, we have 
the following relationships among $a_k$, $c_k$, and $d_k$:
%
\begin{eqnarray}
\rm{Powers} &          \rm{Relation\ Among\ Coefficients} & \notag \\
(z-z_0)^0   &                           c_0 &= a_0 \notag \\
(z-z_0)^1   &                     c_1-a_0d_1 &= a_1 \notag \\
(z-z_0)^2   &              c_2-a_1d_1-a_0d_2 &= a_2  \\
(z-z_0)^3   &        c_3-a_2d_1-a_1d_2-a_0d_3 &= a_3 \notag \\
\cdots \cdots       &  \cdots \cdots \cdots \cdots  & \notag \\
(z-z_0)^k   & \, \, \, \, \, \, c_k-a_{k-1}d_1-a_{k-2}d_2-\cdots-a_0d_k &= a_k \notag 
\label{relatecoeff}
\end{eqnarray}
where $c_k=0$ for $k>M$ and $d_k=0$ for $k>N$. In a matrix language we have 


\normalsize
\begin{eqnarray}
\left(
\begin{array}{cccccccccc}
1      & 0      & 0      & \ldots & 0       & -a_0       & 0         & 0         & \ldots  & 0         \\
0      & 1      & 0      & \ldots & 0       & -a_1       & -a_0       & 0         & \ldots  & 0         \\
0      & 0      & 1      & \ldots & 0       & -a_2       & -a_1       & -a_0       & \ldots  & 0         \\
\vdots & \vdots & \vdots & \ddots & \vdots  & \vdots    & \vdots    & \vdots    & \ddots  & \vdots    \\
0      & 0      & 0      & \ldots & 1       & -a_{M-1}   & -a_{M-2}   & -a_{M-3}   & \ldots  & -a_{M-N}   \\
0      & 0      & 0      & \ldots & 0       & -a_{M}     & -a_{M-1}   & -a_{M-2}   & \ldots  & -a_{M-N+1} \\
0      & 0      & 0      & \ldots & 0       & -a_{M+1}   & -a_{M}     & -a_{M-1}   & \ldots  & -a_{M-N+2} \\
0      & 0      & 0      & \ldots & 0       & -a_{M+2}   & -a_{M+1}   & -a_{M}     & \ldots  & -a_{M-N+3} \\
\vdots & \vdots & \vdots & \ddots & \vdots  & \vdots    & \vdots    & \vdots    & \ddots  & \vdots    \\
0      &  0     & 0      & \ldots & 0       & -a_{M+N-1} & -a_{M+N-2} & -a_{M+N-3} & \ldots  & -a_{M}
\end{array}
\right)
\left(
\begin{array}{c}
c_1 \\
c_2 \\
c_3 \\
\vdots \\
c_M \\
d_1 \\
d_2 \\
d_3 \\
\vdots \\
d_N
\end{array}
\right)
=
\left(
\begin{array}{c}
a_1 \\
a_2 \\
a_3 \\
\vdots \\
a_{M} \\
a_{M+1} \\
a_{M+2} \\
a_{M+3} \\
\vdots \\
a_{M+N}
\end{array}
\right)\notag
\\ 
\label{eq:cmplxArrayEqn}
\end{eqnarray}

\large

\noindent
where $a_k=0$ for $k<0$. By inverting this matrix, we can determine the 
coefficients $c_j$ and $d_k$ $(j=1, \cdots,M {\rm\ and\ } k=1,\cdots,N)$.
Note: The inversion of this kind of matrices is easily done by computers.
(Cf. IBM 360 Scientific Subroutine Package (SSP).)
\\

\noindent
 (PADE 1)\\
\hspace*{1cm}The Taylor series of $w(z)$ around the origin is \cite{Abramowitz}  
\begin{equation}
w(z) = \sum_{j=0}^{\infty}a_jz^j
     = \sum_{j=0}^{\infty}\frac{(iz)^j}{\Gamma(j/2+1)}.
\label{eq:Pade.6}
\end{equation}
Let
\begin{equation}
u = iz = -ZI + iZ\!R,\quad\hbox{where\ } z = Z\!R + iZI.
\label{eq:Pade.7}
\end{equation}
Then
\begin{equation}
w(z) \overset{\rm Def.}{\ =\ } G(u) = \sum_{j=0}^{\infty}\,\frac{u^j}{\Gamma(j/2+1)}.
\label{eq:Pade.8}
\end{equation}
We shall apply the Pad\'e approximation to $G(u)$.  Considering the 
behavior of $w(z)$
\begin{equation}
w(z) \rightarrow 0 {\rm\ as\ } |z| \rightarrow \infty 
\notag
\end{equation}
for those $z$ such that $|Z\!R| > |ZI|$, we take
%
\begin{equation}
 M=6 {\rm\ and\ } N=7. 
\notag
\end{equation}
By inverting the matrix~(\ref{eq:cmplxArrayEqn}), 
we obtain, up to nine significant figures, 
\begin{eqnarray}
&c_0 = 1 \, \, ({\rm Cf.}\, (7)) \, \, \, &d_1 = -2.38485635 \notag \\
&c_1 = -1.25647718 \, \, \, &d_2 = 2.51608137 \notag \\
&c_2 = 8.25059158 \times 10^{-1} \, \, \, &d_3 = -1.52579040 \notag \\
&c_3 = -3.19300157 \times 10^{-1} \, \, \, &d_4 = 5.75922693 \times 10^{-1} \\
&c_4 = 7.63191605 \times 10^{-2} \, \, \, &d_5 = -1.35740709 \times 10^{-1} \notag \\
&c_5 = -1.04697938 \times 10^{-2} \, \, \, &d_6 = 1.85678083 \times 10^{-2} \notag \\
&c_6 = 6.44878652 \times 10^{-4} \, \, \, &d_7 = -1.14243694 \times 10^{-3} \notag
\label{coeff1}
\end{eqnarray}
Hence, the approximation of $w(z)$ near the origin is, by (\ref{eq:Pade.1}), 
\begin{equation}
w(z) = G(u) 
  \simeq
\frac{1 + c_1u + c_2u^2 + c_3u^3 +c_4u^4 + c_5u^5 + c_6u^6}
     {1 + d_1u + d_2u^2 + d_3u^3 +d_4u^4 + d_5u^5 + d_6u^6 + d_7u^7},
\label{eq:Pade.10}
\end{equation}
%
where $u=-ZI+iZR$, and the coefficients $c_k$ and $d_k$ are given by (12).
\\

\noindent
 (PADE 2)\\
\hspace*{1cm}Since the approximation PADE 1 behaves rather poorly along the real axis
right around $z=3$ (Cf. Table 2), we need a Pad\'e approximation around 
$z=3$. The Taylor series of $w(z)$ at $z=3$ is
\begin{equation}
w(z) = \sum_{j=0}^{\infty}\,a_j (z-3)^j,
\label{eq:Pade.11}
\end{equation}
where
\begin{equation}
a_j = \frac{w^{(j)}(3)}{j!}.
\label{eq:Pade.12}
\end{equation}
The derivatives $w^{(j)}(3)$ can be expressed in terms of $w(3)$ by use of
the relations \cite{Abramowitz}
\begin{eqnarray}
w^{(j+2)}(z) &+& 2zw^{(j+1)}(z) + 2(j+1)w^{(j)}(z)=0,\quad (j=0,1,2,\cdots) \notag\\
w^{(0)}(z) &=& w(z),\quad w'(z)=-2zw(z) + \displaystyle{\frac{2i}{\sqrt\pi}}.
\label{eq:Pade.13}
\end{eqnarray}
On the other hand, the value of $w(3)$ is, by (\ref{eq:w_z}), 
\begin{equation}
w(3) = e^{-9} + \frac{2i}{\sqrt\pi}\,e^{-9}\,\int_0^3\,e^{u^2}\,du.
\label{eq:Pade.14}
\end{equation}
By using Table~2 in Rosser \cite{Rosser} for the value of the second term, we 
have $w(3)$ up to nine significant figures:
\begin{equation}
w(3) = 1.23409804\times10^{-4} + i2.01157318\times10^{-1}.
\label{eq:Pade.15}
\end{equation}
This time we choose 
\begin{equation}
 M=3 {\rm\ and\ } N=4. 
\notag
\end{equation}
By inverting the matrix~(\ref{eq:cmplxArrayEqn}),
we obtain, up to nine significant figures, 
\begin{eqnarray}
c_0 &=& 1.23409804 \times 10^{-4} + i2.01157318 \times 10^{-1} \, \, ({\rm Cf.}\, (7)) \notag \\
c_1 &=& 2.33746715 \times 10^{-1} + i1.61133338 \times 10^{-1} \notag \\
c_2 &=& 1.25689814 \times 10^{-1} - i4.04227250 \times 10^{-2} \notag \\
c_3 &=& 8.92089179 \times 10^{-3} - i1.81293213 \times 10^{-2} \\
d_1 &=& 1.19230984 - i1.16495901 \notag \\
d_2 &=& 8.94015450 \times 10^{-2} - i1.07372867 \notag \\
d_3 &=& -1.68547429 \times 10^{-1} - i2.70096451 \times 10^{-1} \notag \\
d_4 &=& -3.20997564 \times 10^{-2} - i1.58578639 \times 10^{-2} \notag 
\label{coeff3}
\end{eqnarray}
Hence, the approximation of $w(z)$ near $z=3$ is, by (\ref{eq:Pade.1}), 
\begin{equation}
w(z) 
  \simeq
\frac{c_0 + c_1z + c_2z^2 + c_3z^3}
     {1 + d_1z + d_2z^2 + d_3z^3 +d_4z^4},
\label{eq:Pade.20}
\end{equation}
%
where the coefficients $c_k$ and $d_k$ are given by (19).
\\

\subsection{Asymptopic Expression}
Away from the origin and $z=3$ we can use the asymptotic expression 
for $w(z)$ given by Faddeyeva and Terent'ev (Eqn.~(10)) \cite{Faddeyeva}.
The formula is
%
\begin{equation}
w(z) 
\simeq
\sum_{k=1}^n\,\frac{i\lambda_k^{(n)}}{\pi(z-x_k^{(n)})}
= 
\sum_{k=1}^n\,\frac{ia_k^{(n)}}{z-x_k^{(n)}}, \quad
a_k^{(n)} = \frac{\lambda_k^{(n)}}{\pi},
\label{eq:Asymp.1}
\end{equation}
%
where $x_k^{(n)}$ are the roots of Hermite polynomials and $\lambda_k^{(n)}$
are the corresponding coefficients (and $n$ is an integer related to the
accuracy of the approximation). The values of $x_k^{(n)}$ and $\lambda_k^{(n)}$
are found in Greenwood and Miller \cite{Greenwood}.  By choosing $n=10$, we 
have an asymptotic expression of $w(z)$ as 
%
\begin{eqnarray}
w(z) 
&\simeq&
  \frac{ia_1}{z-x_1} + \frac{ia_1}{z+x_1} 
+ \frac{ia_2}{z-x_2} + \frac{ia_2}{z+x_2} 
+ \frac{ia_3}{z-x_3} + \frac{ia_3}{z+x_3} \notag \\
&+& \frac{ia_4}{z-x_4} + \frac{ia_4}{z+x_4} 
+ \frac{ia_5}{z-x_5} + \frac{ia_5}{z+x_5}, 
\label{eq:Asymp.2}
\end{eqnarray}
%
where the constants are, up to nine or ten significant figures,
\begin{eqnarray}
&a_1 = 1.94443615 \times 10^{-1} \, \, \, &x_1 = 3.42901327 \times 10^{-1} \notag \\
&a_2 = 7.64384940 \times 10^{-2} \, \, \, &x_2 = 1.036610830 \notag \\
&a_3 = 1.07825546 \times 10^{-2} \, \, \, &x_3 = 1.756683649 \\
&a_4 = 4.27695730 \times 10^{-4} \, \, \, &x_4 = 2.532731674 \notag \\
&a_5 = 2.43202531 \times 10^{-6} \, \, \, &x_5 = 3.436159119 \notag
\label{coeff4}
\end{eqnarray}

\subsection{Regions of Validity of the Three Approximations}
The regions of validity of the three approximations are illustrated in
Figures 1 and 2, which will be explained below in detail.

In order to check our approximations we used the tables of $w(z)$ by
Faddeyeva and Terent'ev \cite{Faddeyeva}. The tables give six-place values of $w(z)$ for the
square $0 \le Z\!R \le 3$, $0 \le ZI \le 3$ with tabular step of 0.02 for each of the
variables and six-place values of $w(z)$ for the range $3 \le Z\!R \le 5$, $0 \le ZI \le 3$ and
$0 \le Z\!R \le 5$, $3 \le ZI \le 5$ with tabular step of 0.1 for each of the variables.
We also used, as a reference in computing, the formulae (Cf. Abramowitz and
Stegun, Eqn. 7.1.26 and 7.1.29)
\begin{equation}
 \textnormal{erf}(Z\!R)\simeq 1-(a_1t+a_2t^2+a_3t^3+a_4t^4+a_5t^5)e^{-Z\!R^2} , 
\label{eq:erfZR}
\end{equation}
where $t=\displaystyle{\frac{1}{1+pZ\!R}}$ and $p, a_1, a_2, a_3, a_4, a_5$ are real constants and
%
%
\begin{eqnarray}
 \textnormal{erf}(Z\!R+i\,ZI) &\simeq& \textnormal{erf}(Z\!R) + 
\frac{e^{-Z\!R^2}}{2\pi Z\!R}\{(1-\textnormal{cos}(2Z\!RZI))+i\, \textnormal{sin}(2Z\!RZI)\} + \notag \\
 &~&\frac{2e^{-Z\!R^2}}{\pi}\sum_{n=1}^{\infty}\frac{e^{-n^2/4}}{n^2+4Z\!R^2}\{f_n(Z\!R,ZI)+i\,g_n(Z\!R,ZI) \} , \notag
\\
\label{eq:erfZRZI}
\end{eqnarray}
where
\begin{eqnarray}
&&f_n(Z\!R,ZI)=2Z\!R-2Z\!R\textnormal{cosh}(nZI)\textnormal{cos}(2Z\!RZI)+n\textnormal{sinh}(nZI)\textnormal{sin}(2Z\!RZI) , \notag \\
%
&&{\rm and} \notag \\
%
&&g_n(Z\!R,ZI)=2Z\!R\textnormal{cosh}(nZI)\textnormal{sin}(2Z\!RZI)+n\textnormal{sinh}(nZI)\textnormal{cos}(2Z\!RZI) . \notag 
\end{eqnarray}
These formulae allow us to calculate the percent error of the approximations,
i.e., $100\, \times$ (Approximation - Exact Value) / (Exact Value), by computer 
(Cf. Program 6). Unfortunately, as we can tell from Table 1, Program 6 which
evaluates $w(z)$ through (\ref{eq:erfZR}) and (\ref{eq:erfZRZI}) does not give quite accurate values,
especially for those regions where $Z\!R$ is small and $ZI$ is large simultaneously.
Thence, the percent errors given in Table 2 through Table 6 are not very
reliable in those 
 ``bad'' regions. In other words our rational approximations are
normally more accurate than the reference formula and hence the listed errors
are over-estimated.

\bigskip
\noindent
(PADE 1)\\
\hspace*{1cm}(The region of validity of PADE 1 is illustrated in Figure 1.)\\
We computed PADE 1, i.e., Eqn. (\ref{eq:Pade.10}) (Cf. Program 3), up to nine significant
places in the range $0 \le Z\!R \le 5$, $0 \le ZI \le 5$ with step of 0.1 and checked
the results against the tables by Faddeyeva and Terent'ev. The agreement was
excellent except along the real axis with $Z\!R$ large; even at $Z\!R=ZI=5$, the real
part of PADE 1 agreed with the table up to six places, maximum accuracy of the
table, and the imaginary part of PADE 1 agreed with that of the table up to five
places. On the real axis, we found percent errors of $\sim$1.3\% at $Z\!R=2.9$ and 
$\sim$2.9\%
at $Z\!R=3.0$ for the $real$ part of $w(z)$, and even larger error for larger $Z\!R$ (Cf.
Table 2). But we note that PADE 1 is very accurate for $ZI=0.1$ (even with $Z\!R=5$).
The breakdown does not occur unless $ZI$ is very small ($\sim$0.01 or smaller). We
also note that the imaginary part of PADE 1 is very accurate even in this area.

\bigskip
\noindent
(PADE 2)\\
\hspace*{1cm}(The region of validity of PADE 2 is illustrated in Figures 1 and 2.)\\
We computed PADE 2, i.e., Eqn. (\ref{eq:Pade.20}) (Cf. Program 4), up to nine significant 
places in exactly the same region as in PADE 1 and checked the results
against the tables by Faddeyeva and Terent'ev. The agreement was good (not as
good as in PADE 1) even away from the point $z=3$. The percent errors were
\noindent
much less than 1\% in most of the region except for the points along the real
axis with $Z\!R$ large and the points near the imaginary axis (e.g., at the origin,
$\sim$2\% error and at $Z\!R=4.0$, $ZI=0$, $\sim$15\% error) (Cf. Table 3; also see Figure 1 for
the errors near the real axis). We note that the breakdown near the real axis is
abrupt just as for PADE 1, i.e., the approximation is good until $ZI$ gets very
small ($\sim$0.01 or smaller). Again the imaginary part of PADE 2 is very accurate
even on the real axis.

\bigskip
\noindent
(ASYMP)\\
\hspace*{1cm}(The region of validity of ASYMP, i.e., the.asymptotic formula (\ref{eq:Asymp.2}) \\
\hspace*{1cm} (Cf. Program 5), is illustrated in Figures 1 and 2.)\\
Exactly the same procedures as for PADE 1 and PADE 2 were followed. The
approximation is excellent for $ZI$ large enough ( $\gtrsim$ 1.0 ) or $Z\!R$ large. But again
the real part is a poor approximation on the real axis (Cf. Table 4). In fact,
Eqn. (22) implies that the real part of $w(z)$ is zero on the real axis, which
is a 100\% error. Hence, even though ASYMP becomes a better approximation as
$Z\!R$ gets larger, the valid region of the real part of ASYMP never reaches the
real axis (e.g., Figure 2 implies that ASYMP is good for $ZI \sim 0.002$ at 
$Z\!R \sim 4.2$).
Again the imaginary part of ASYMP is very accurate even in this region.
To overcome the difficulty we expanded $w(z)$ in powers of $ZI$ and kept only the
first power in $ZI$ as follows. For $ZI \ll 1$ and $Z\!RZI \ll 1$ we have, keeping only
the first power of $ZI$ in (3),
\begin{equation*}
 w(z)\simeq e^{-Z\!R^2}(1-i\,2Z\!RZI)\left(1+\frac{2i}{\sqrt{\pi}}\int\displaylimits_0^{Z\!R}e^{u^2}du-\frac{2}{\sqrt{\pi}}e^{{Z\!R}^2}ZI\right) . 
\label{eq:wzltlt1}
\end{equation*}
%
Thus, the real part is, for $Z\!RZI \ll 1$ and $ZI \ll 1$,
\begin{equation}
 \Re w(z)=e^{-Z\!R^2}+2\left\{Z\!R \Im w(Z\!R+i0)-\frac{1}{\sqrt{\pi}}\right\}ZI . 
\label{eq:rewzltlt1}
\end{equation}
%
Note: The formula (26) is plausible because the imaginary part of ASYMP is
very accurate for $Z\!R$ large enough.
The condition for (26) to be valid within 1\% error is
%
\begin{equation}
Z\!RZI\lesssim0.01 . 
\label{eq:region}
\end{equation}
%
We shall discuss this region of validity more in detail in the next subsection.

\subsection{Boundaries of the Valid Regions of the Three Approximations}
\hspace*{1cm} (The reader is again referred to Figures 1 and 2 for illustrations.) \\
Having examined the regions of validity of the three approximations,
our next task is to determine where we should set the boundaries of the three
approximations so that we have minimum possible errors. Given any two of the
three approximations, the idea is to find $ZI$ (or $Z\!R$) for fixed 
$Z\!R$ (or $ZI$) where
we have the {\it least} (or {\it minimum}) {\it discontinuity} between the two 
approximations.
The points of least discontinuity are plotted in Figures 1 and 2. The 
boundaries were set so that they go through as many points of least discontinuity 
as possible.

From the discussions in the previous section we recall that there are bad
points for the real part of $w(z)$ on the real axis inside the PADE 2 region and
the ASYMP region. Since the power expansion formula (26) is a good 
approximation near the real axis (exact on the real axis), we use it there. In Figure
2 we plot the points of least discontinuity both between PADE 2 and the power
expansion and between ASYMP and the power expansion. The boundary between
PADE 2 and the power expansion is fitted by a straight line
\begin{equation}
Z\!RZI=0.0625(Z\!R-3.5) . 
\label{eq:ZRZI2}
\end{equation}
The boundary between ASYMP and the power expansion is fitted by
\begin{equation}
Z\!RZI=\frac{a}{Z\!R-b}+c,\, \, \, \, \, (a,\,b,\,c\,\, \, \, \textnormal{constants}) . 
\label{eq:ZRZI2b}
\end{equation}
Using the three points of least discontinuity, $(Z\!R, Z\!RZI) = (3.8, 0.044)$,
(3.9, 0.0312), and (4.0, 0.022), we find
\begin{equation}
a = 0.04 ,\,  b = 3.29,\,  \textnormal{ and } c=-0.034 . 
\label{eq:abc}
\end{equation}
For $Z\!R > 4.2$ we use the boundary
\begin{equation}
Z\!RZI=0.01 . 
\label{eq:ZRZI3}
\end{equation}
%
\noindent
To sum up:\\\\
ASYMP is modified so that it calculates the power expansion formula (26)\\\\
if $Z\!R < 4.2$ and $Z\!RZI < \displaystyle{\frac{0.04}{Z\!R-3.29}}- 0.034$ , \, \, or\\\\
if $Z\!R \ge 4.2$ and $Z\!RZI < 0.01$ . \\\\
After this modification,\\\\
\hspace*{1cm}for $3.5 \le Z\!R < 4.1$\\\\
\hspace*{2cm}use ASYMP if $Z\!RZI < 0.0625(Z\!R-3.5)$ , \\\\
\hspace*{2cm}use PADE 2 if $Z\!RZI \ge 0.0625(Z\!R-3.5)$ , \\\\
\hspace*{1cm} for $Z\!R \ge 4.1$\\\\
\hspace*{2cm}use ASYMP .

\subsection{Electric Field}
Once we have the function $w(z)$, we can find the electric field by
simply using the formulae~(\ref{eq:E_x}) and (\ref{eq:E_y}).  We set,
for simplicity,
\begin{equation}
\frac{Q}{2\epsilon_0\sqrt\pi} = 1 , 
\label{eq:chargefac}
\end{equation}
in those formulae.

Unfortunately, there is one problem: By symmetry $E_y=0$ for $y=0$.
But we know $\Re w(z)$ is not approximated well near the real axis,
so the two terms in (\ref{eq:E_y}) might not cancel out each other to 
give exactly zero at $y=0$.  This might cause the percent error for $E_y$ 
to be rather large for $y=0$ and $y$ small. To overcome this difficulty
we first set $E_y=0$ if $y=0$ and \emph{linearly interpolate} the values of 
$E_y$ for $y$ very small. That is, for
\begin{equation}
\frac{y}{\sqrt{2(s_x^2-s_y^2)}} < 0.002 ,
\notag
\end{equation}
we set 
\begin{equation}
E_y(x,y)  = \frac{\displaystyle{\frac{y}{\sqrt{2(s_x^2-s_y^2)}}}}{0.002}  
E_y\left(x, 0.002\sqrt{2(s_x^2-s_y^2)}\right).
\label{eq:small-y}
\end{equation}
(Cf. Program 1 and Table 5.) This also serves to guarantee that $E_y(x,y)$
will be continuous between the first and fourth quadrants.

\subsection{Concluding Remarks}
The program FNCTNW calculates $w(z)$ quite accurately. The percent
error in most of the region is $\sim10^{-4}$\% except for the real
part of $w(z)$ near the real axis for certain values of $Z\!R$ 
(near $Z\!R = 2.2,\, 3.5,$ and 4.2) where the percent error could be
at most 0.1\%. 

The program GAFELD likewise calculates the electric field with
the percent error $\sim10^{-4}$\% except for $E_y$ near the real axis 
where the percent error is at most of the order of 0.1\%.

Even though we have rather large percent errors ($\sim$0.1\%) for
$\Re w(z)$ and $E_y$ near the real axis, the \emph{absolute errors}
are small because $\Re w(z)$ and $E_y$ take on small absolute values
there.

We have discussed the accurate evaluation over the entire first 
quadrant. If used in a computer simulation of beam-beam effects, PADE 1
would be called by far the most, as its region of validity more or less 
corresponds to where the particles reside. One may be justified, for the 
sake of simplicity, in regarding PADE 1 as an adequate replacement for 
the true field, but further investigation would be necessary to confirm this.
\\

\leftline{ \bf Acknowledgements }
We would like to thank Professor W. Fuchs in Mathematics Department 
of Cornell University for various useful discussions. 

\bibliographystyle{IEEEtran}

\begin{thebibliography}{1}
\bibitem{Bassetti-typo}
A typographical error in the formula of Bassetti and
  Erskine has been corrected. (The sign of the second term in the exponential
  factor has been reversed.)
\end{thebibliography}

\subsection{Figure Captions}
\begin{enumerate}
\item
Points of least discontinuity among the three approximations and the 
boundaries of separating regions of the three approximations.
The numbers, 2, 3, etc., represent the numbers of decimal places of 
disagreement out of nine significant figures, i.e., 2 means the first seven 
significant
places of agreement and 3 means the first six significant places of
agreement. Those numbers are taken to be the larger one of the two
discontinuities at a point corresponding to the real part and the imaginary part.
The real part and the imaginary part have similar degrees of discontinuity
at each point in most of the region except for those points near the real
axis where the discontinuity of the real part tends to be much bigger than
that of the imaginary part.

\item
Points of least discontinuity between ASYMP (without the power expansion
modification) and the power expansion formula (26) and between PADE 2 and
the power expansion formula.
6, etc. represent the number of decimal places of disagreement between ASYMP
and the power expansion formula.
\textcircled{5}, etc. represent the number of decimal places of disagreement between
PADE 2 and the power expansion formula.
\end{enumerate}

\subsection{Programs}
\noindent
1. GAFELD.FORTRAN\\
2. FNCTNW.FORTRAN\\
3. WPADEl.FORTRAN\\
4. WPADE2.FORTRAN\\
5. WASYMP.FORTRAN\\
6. WEXCT.FORTRAN\\
\\
\hspace*{1cm}To run the computer program for the electric field from the PDP 10 
terminal, we just type\\
.EXE GAFELD, FNCTNW, WPADE1, WPADE2, WASYMP, WEXCT $<$cr$>$

\subsection{Tables}
\noindent
1. WEXCT\\
\hspace*{1cm} The hand-written numbers under some entries represent the 
exact values taken from the tables by Faddeyeva and Terent'ev. Entries 
without any hand-written numbers under them represent those values which 
agree with the tables completely (up to six places).\\
2. PADE 1\\
3. PADE 2\\
4. ASYMP\\
5. Function $w(z)$ ( FNCTNW.FORTRAN )\\
6. Electric Field ( GAFELD.FORTRAN )\\


\clearpage
\begin{minipage}{\textwidth}
\includepdf[pages=1,scale=0.99,pagecommand={}]{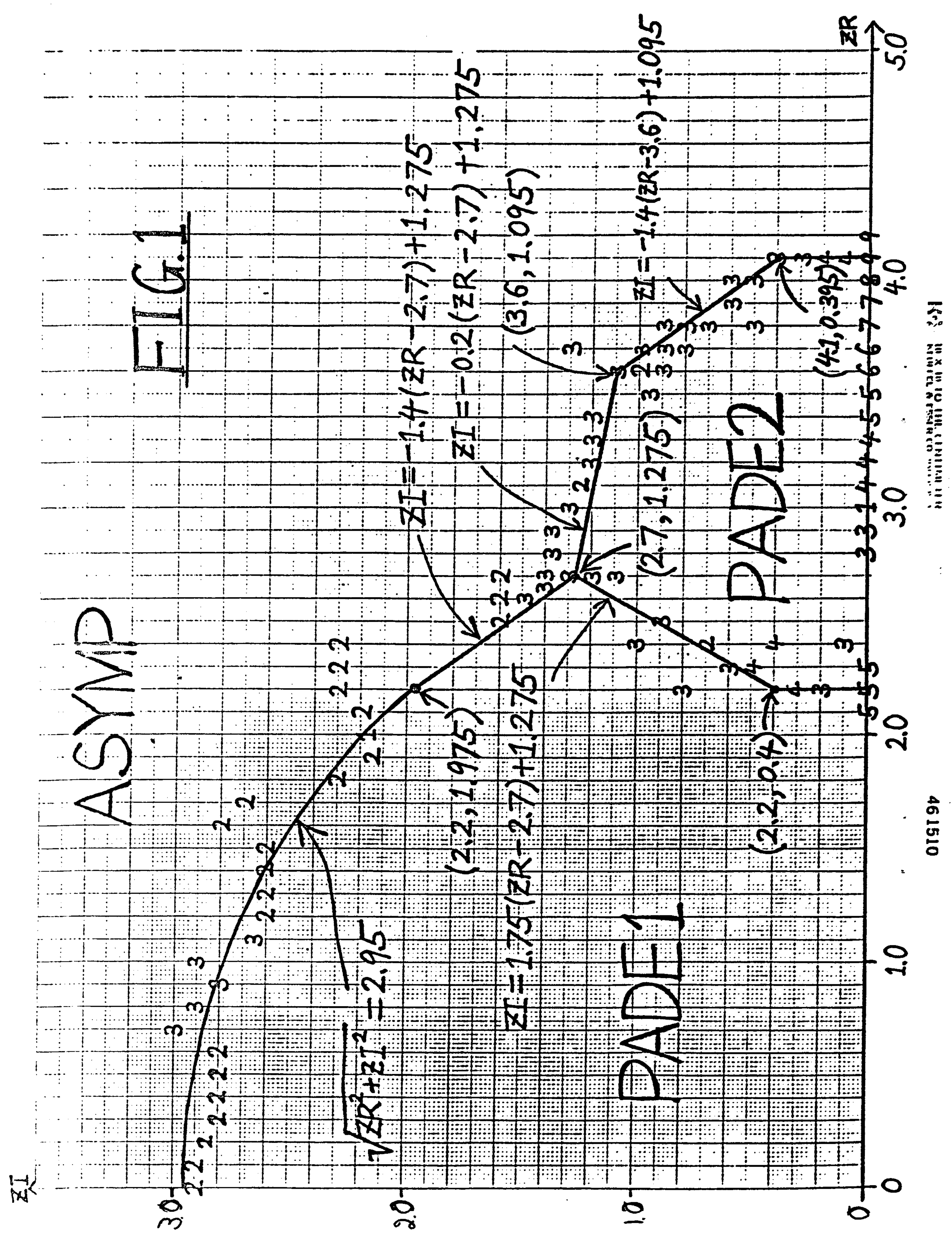}
\end{minipage}

\clearpage
\begin{minipage}{\textwidth}
\includepdf[pages=1,scale=0.95,pagecommand={}]{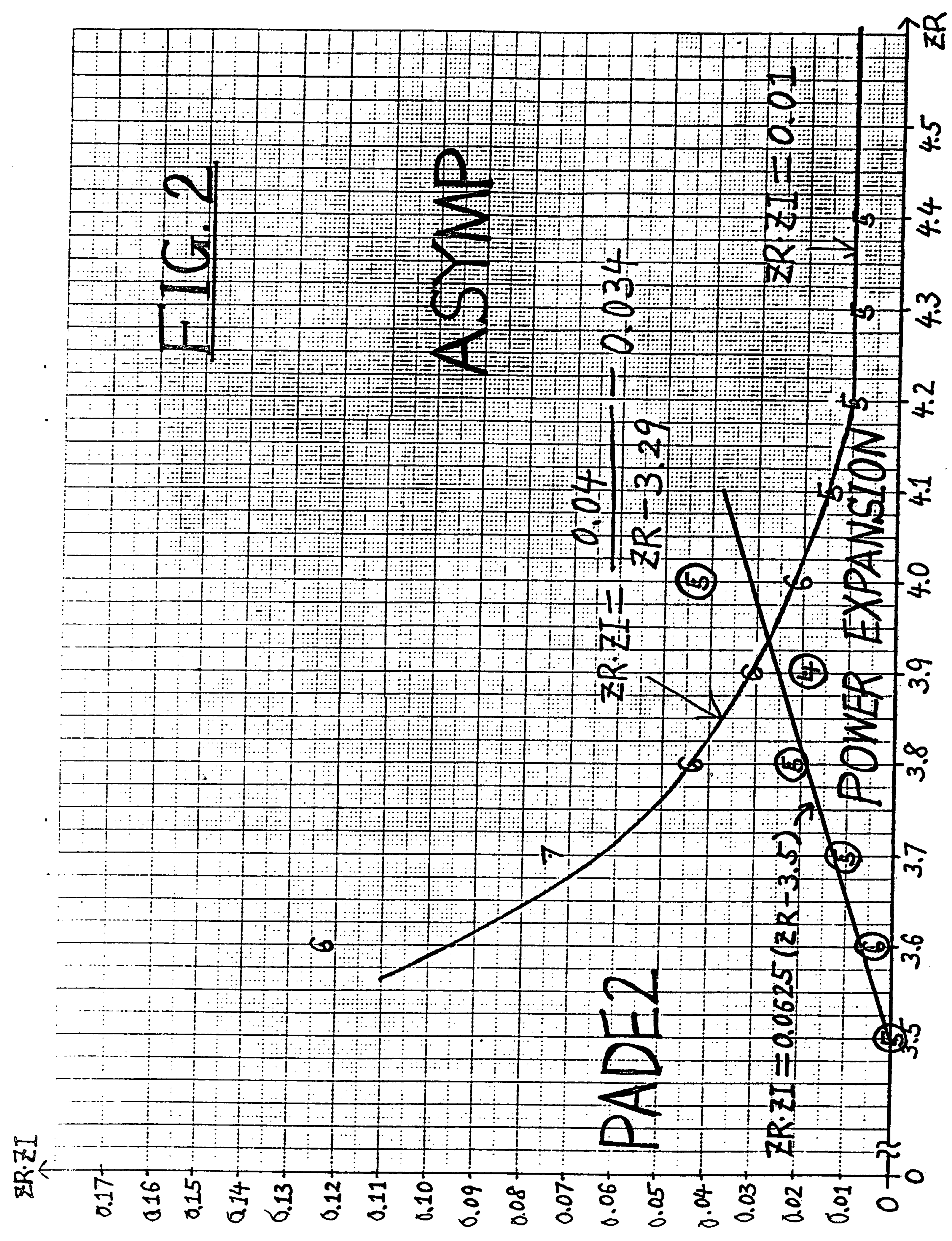} 
\end{minipage}

\clearpage
\begin{minipage}{\textwidth}
\includepdf[pages=1,scale=0.95,pagecommand={}]{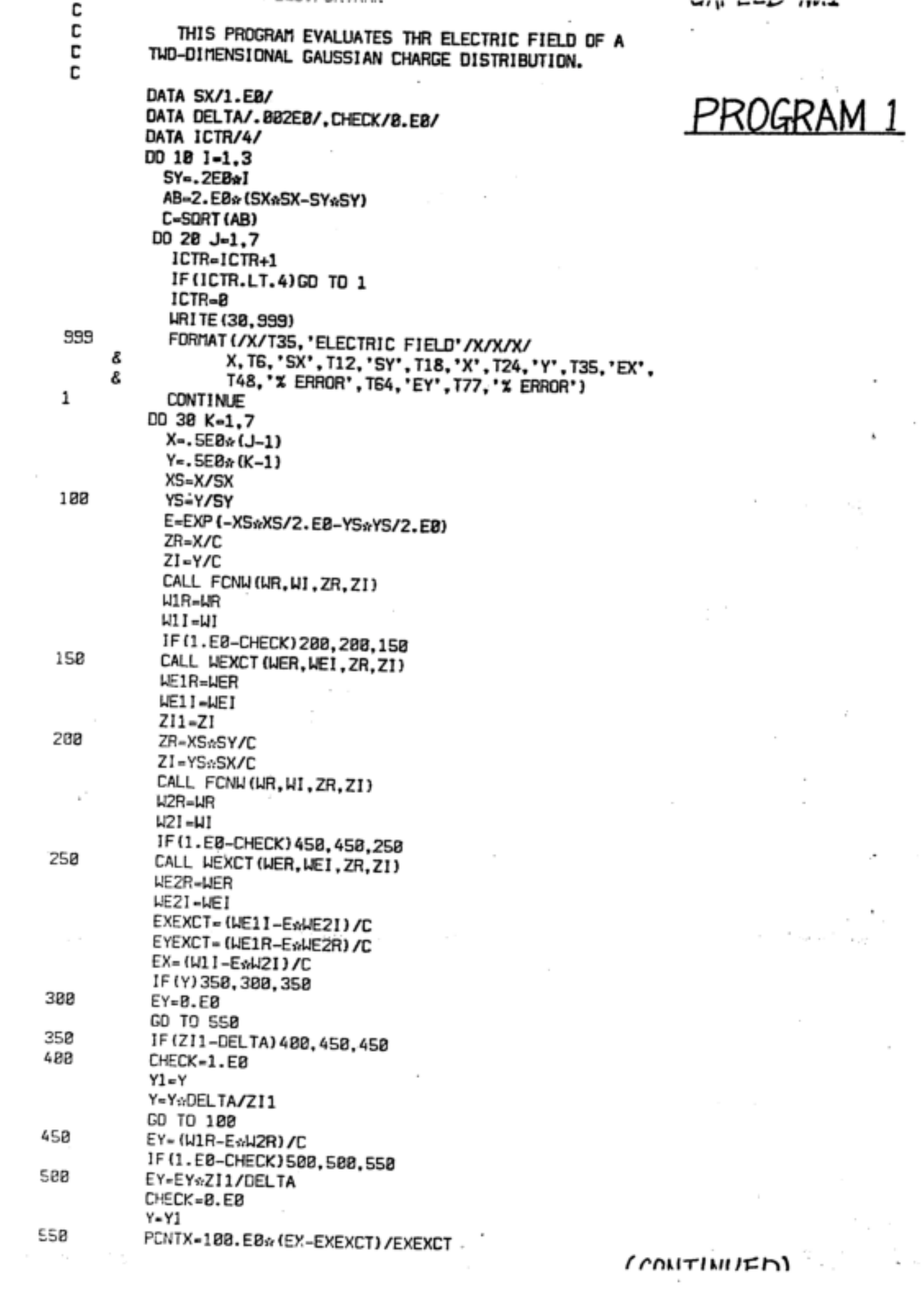} 
\end{minipage}

\clearpage
\begin{minipage}{\textwidth}
\includepdf[pages=1,scale=0.95,pagecommand={}]{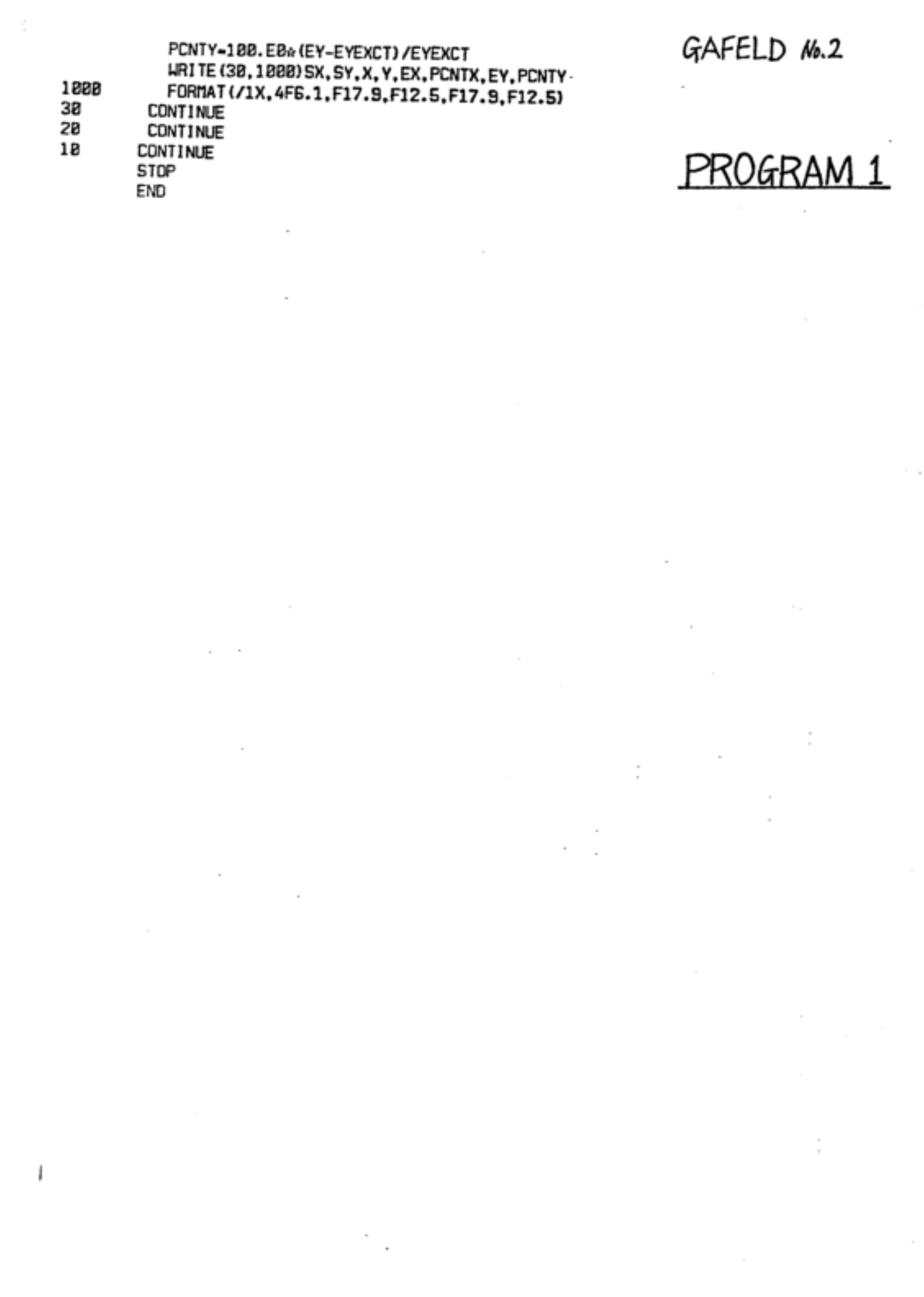} 
\end{minipage}

\clearpage
\begin{minipage}{\textwidth}
\includepdf[pages=1,scale=0.95,pagecommand={}]{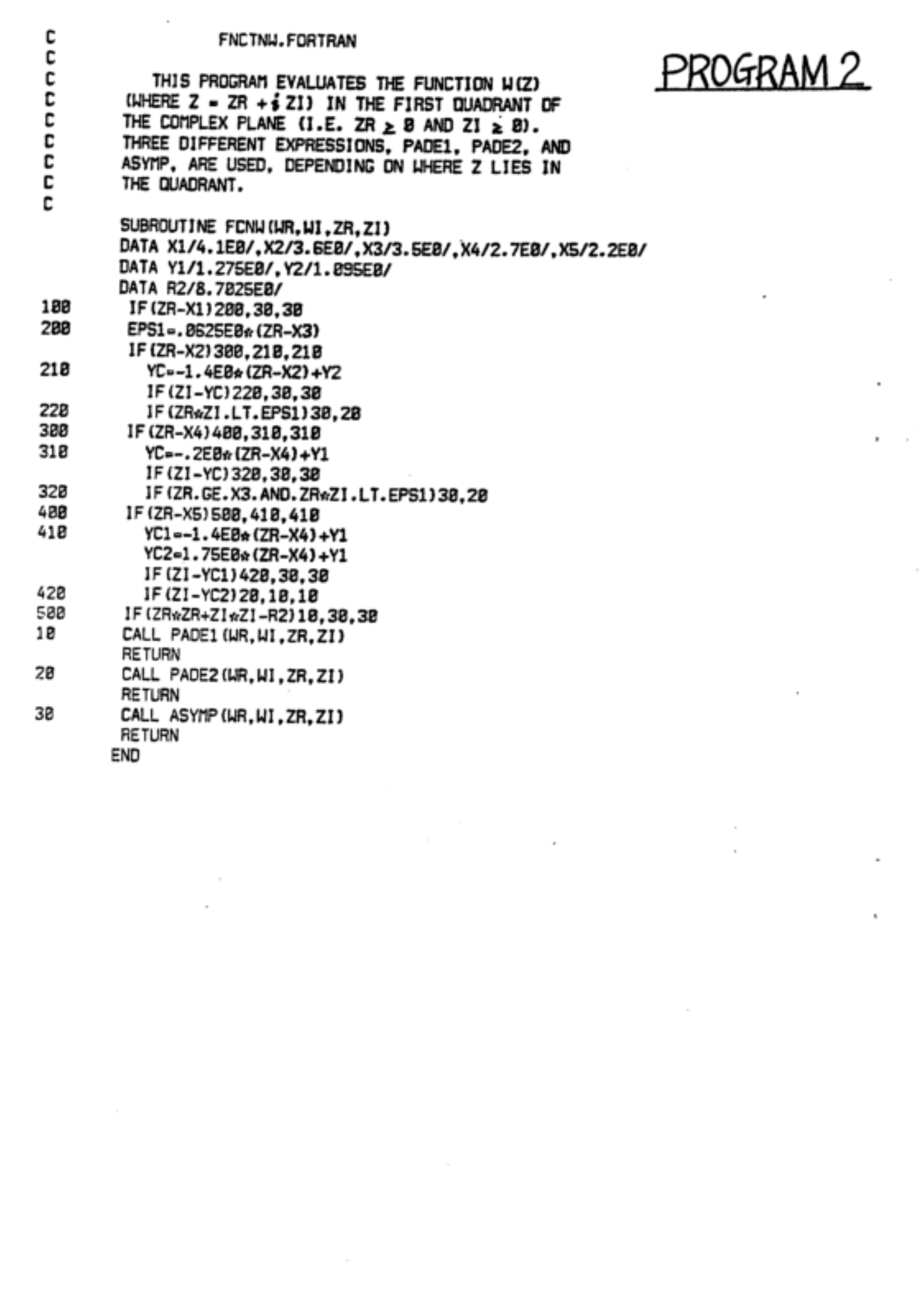} 
\end{minipage}

\clearpage
\begin{minipage}{\textwidth}
\includepdf[pages=1,scale=0.95,pagecommand={}]{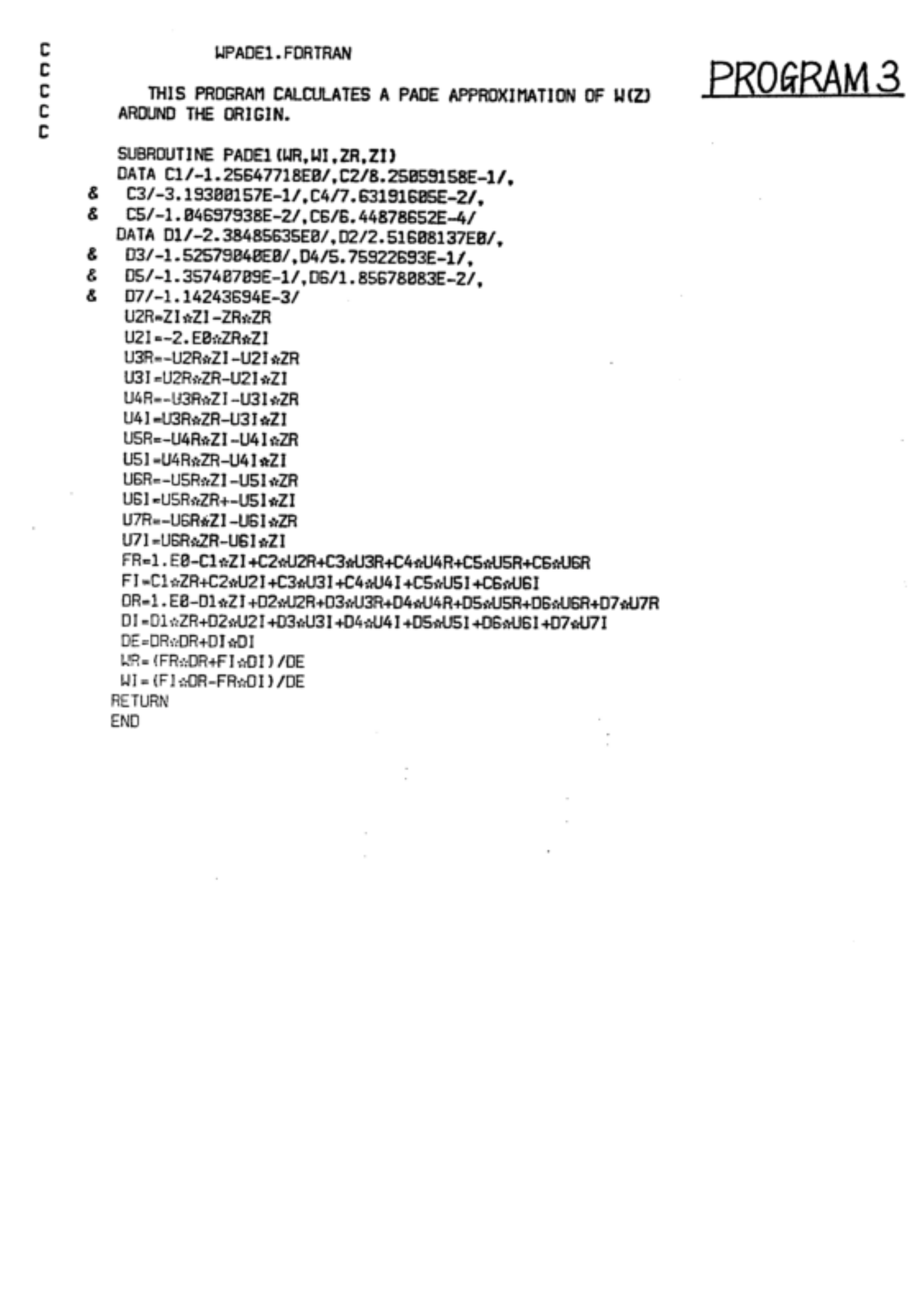} 
\end{minipage}

\clearpage
\begin{minipage}{\textwidth}
\includepdf[pages=1,scale=0.95,pagecommand={}]{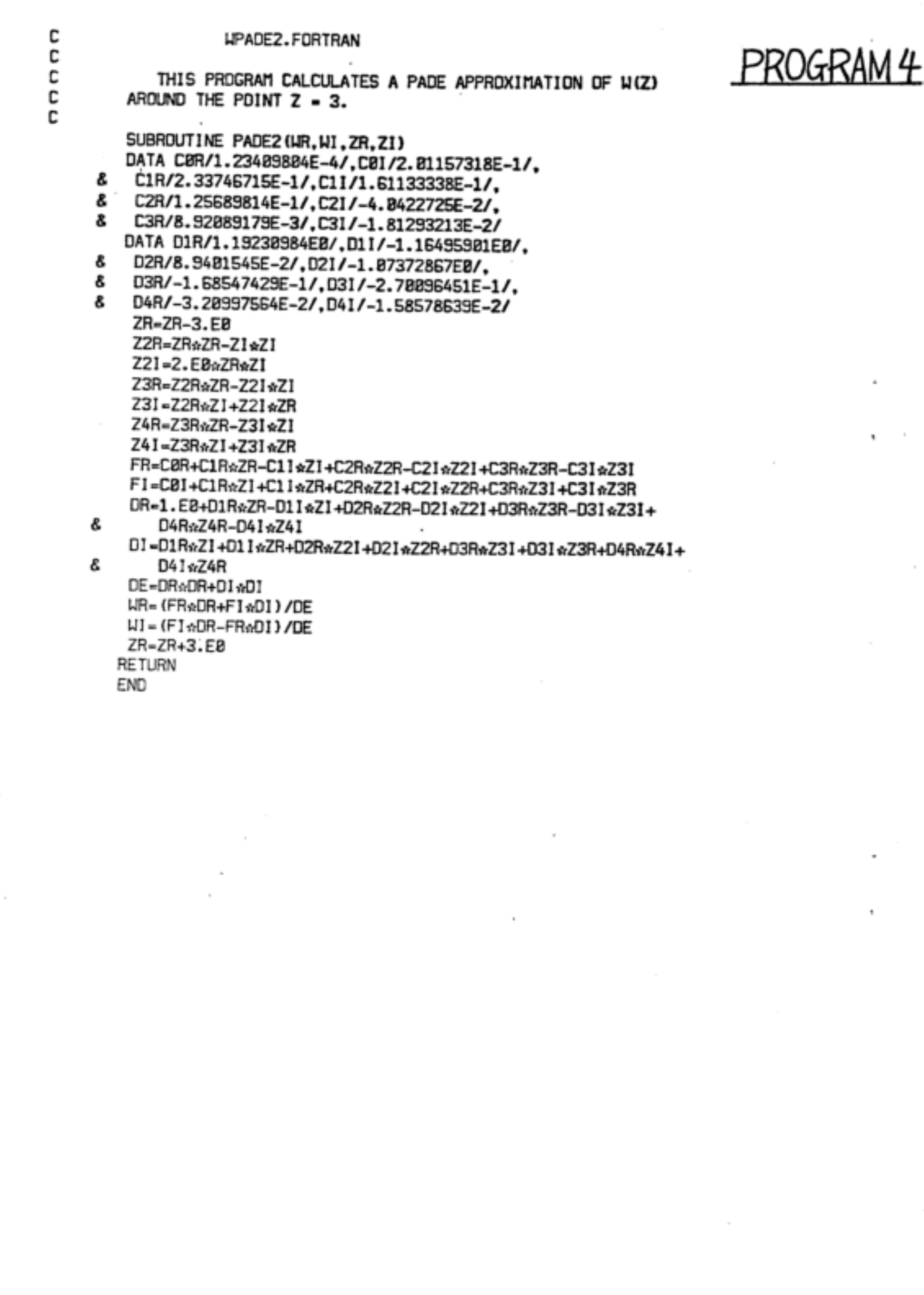} 
\end{minipage}

\clearpage
\begin{minipage}{\textwidth}
\includepdf[pages=1,scale=0.95,pagecommand={}]{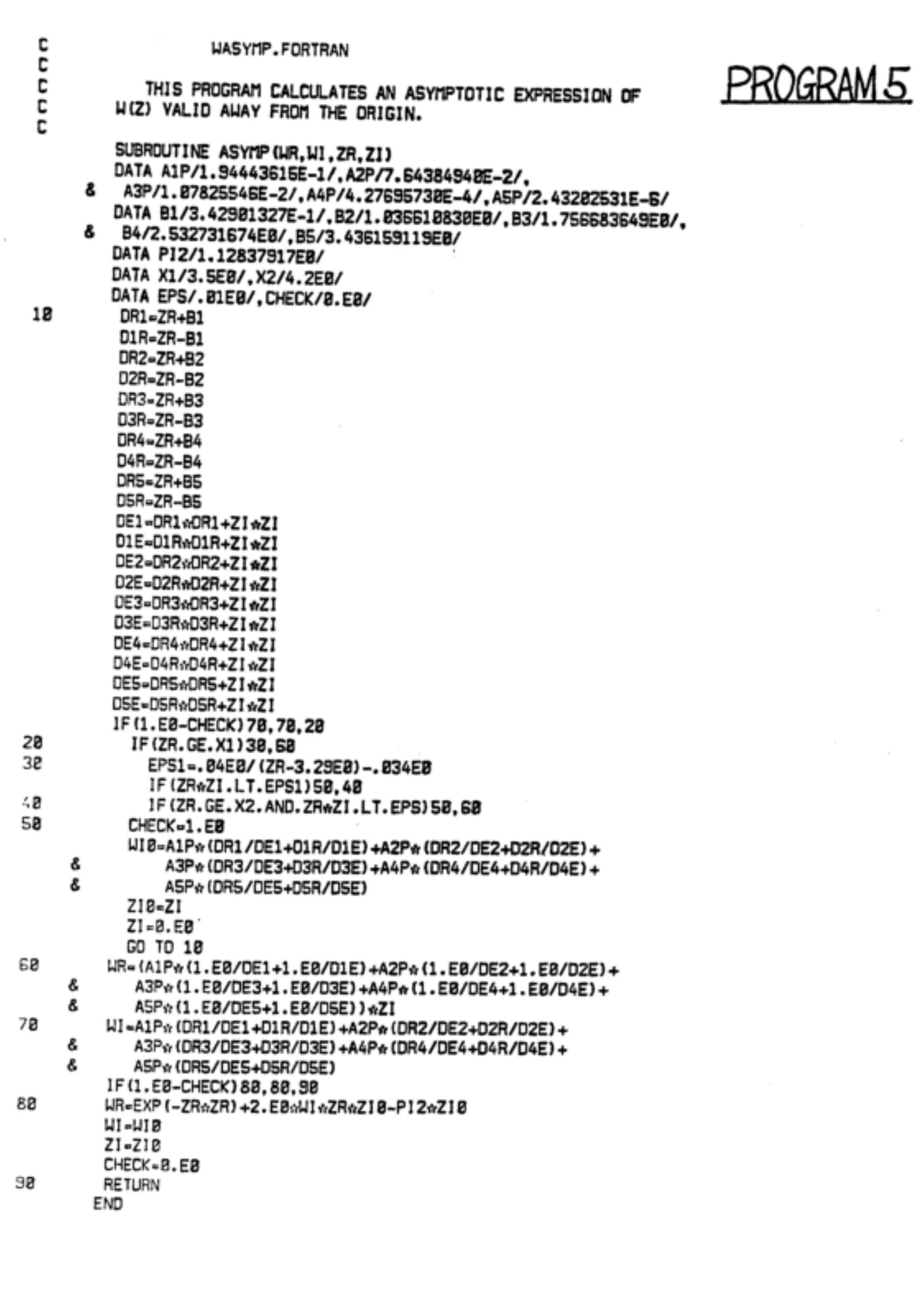} 
\end{minipage}

\clearpage
\begin{minipage}{\textwidth}
\includepdf[pages=1,scale=0.95,pagecommand={}]{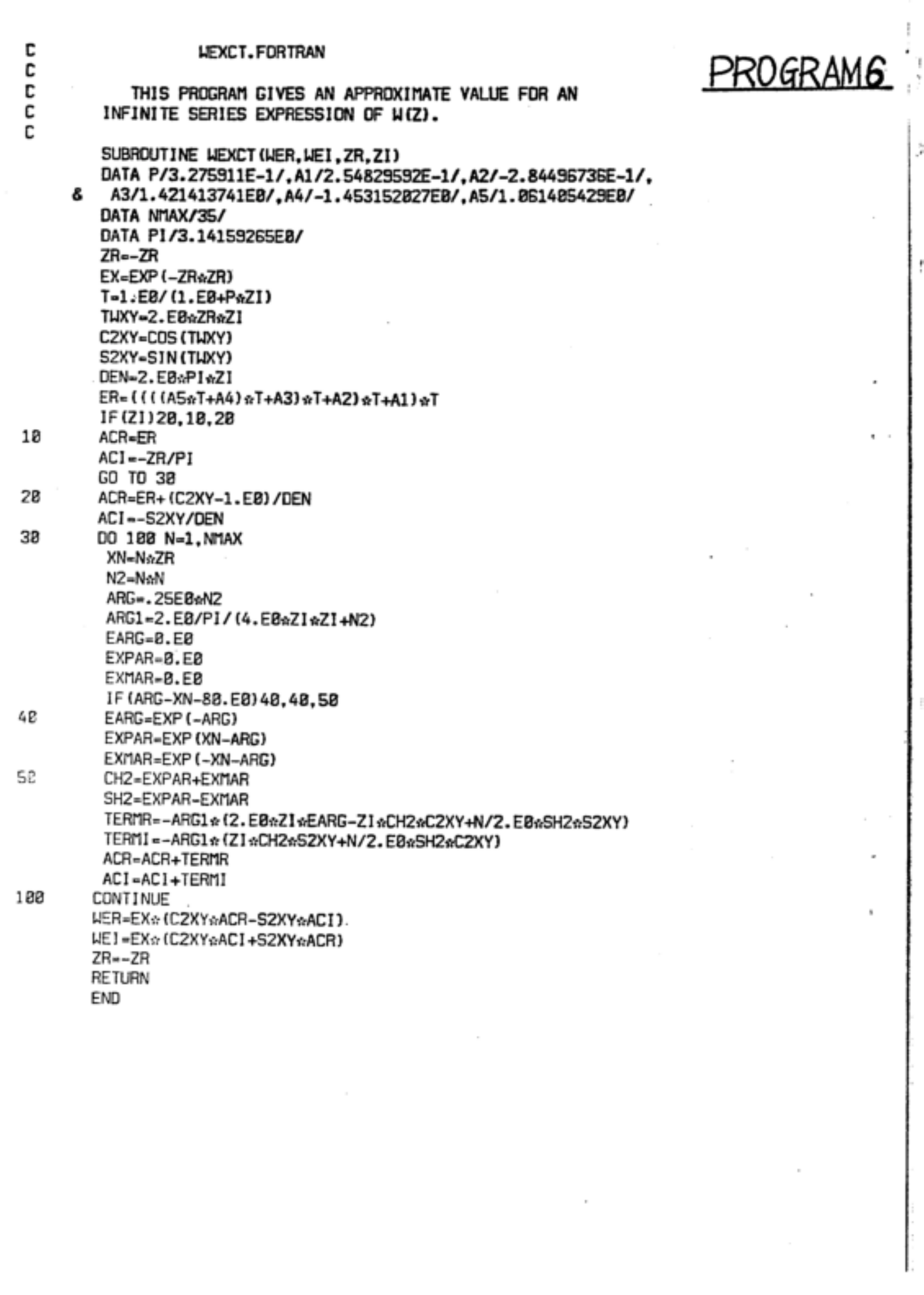} 
\end{minipage}

\clearpage
\begin{minipage}{\textwidth}
\includepdf[pages=1,scale=0.95,pagecommand={}]{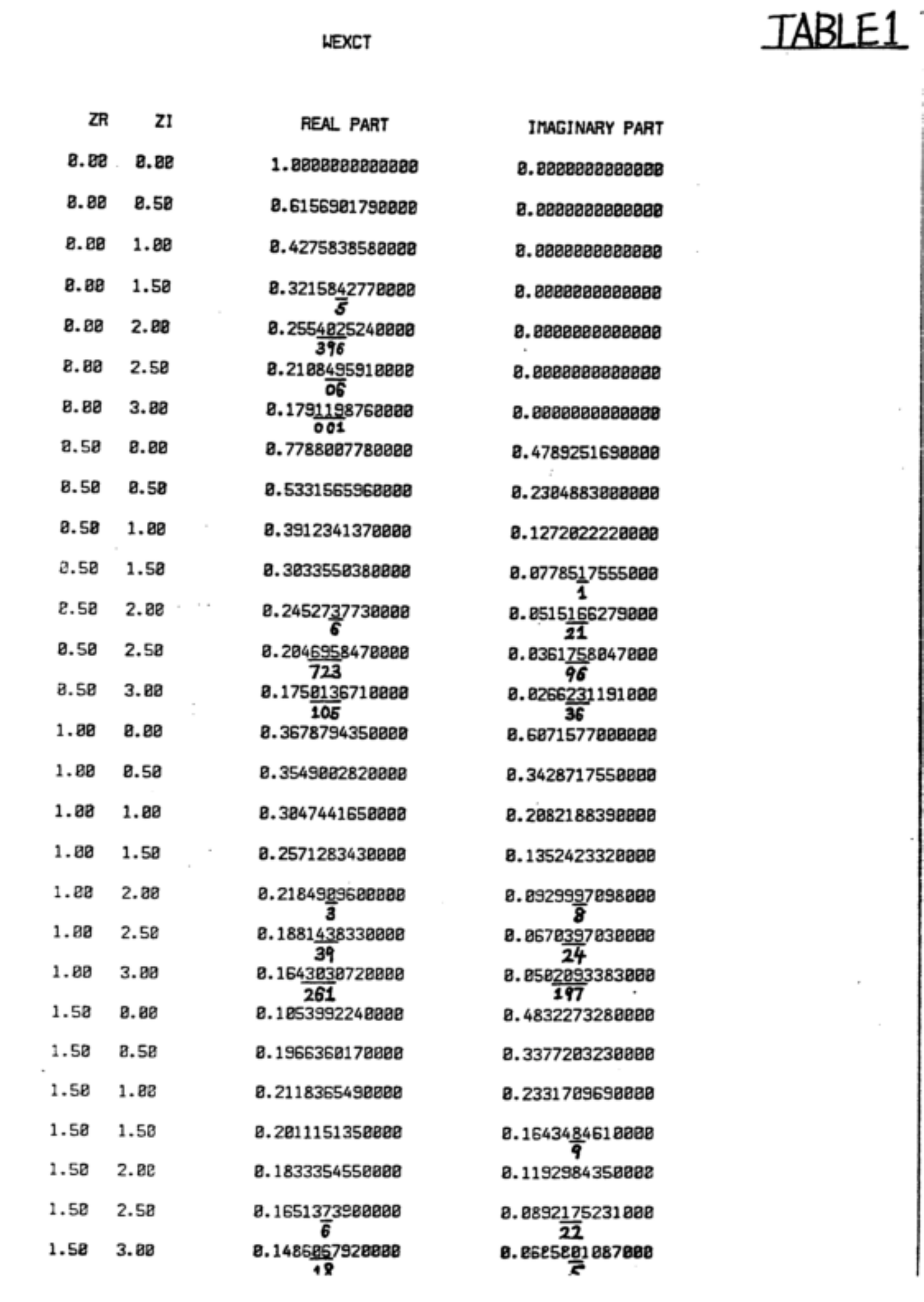} 
\end{minipage}

\clearpage
\begin{minipage}{\textwidth}
\includepdf[pages=1,scale=0.95,pagecommand={}]{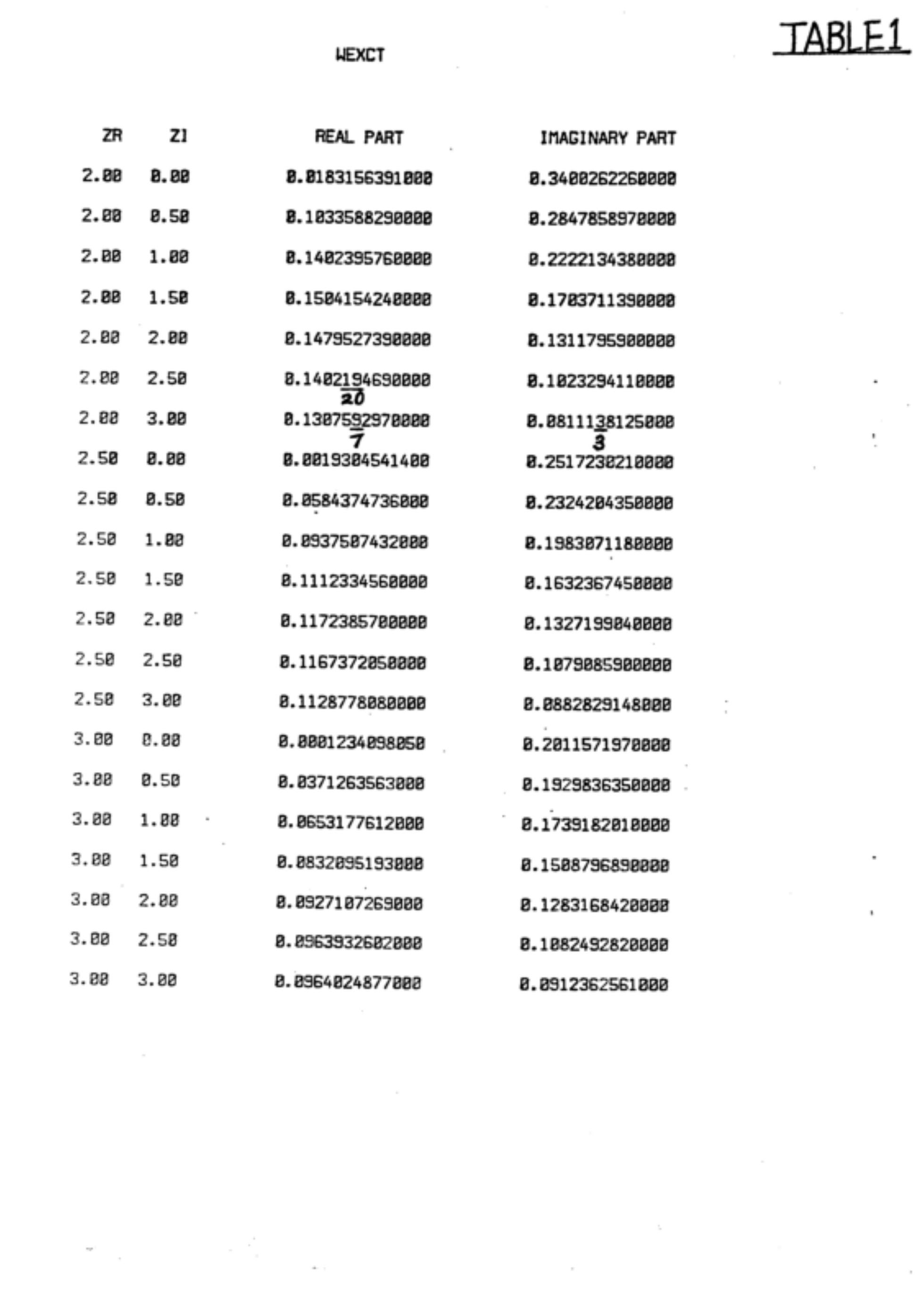} 
\end{minipage}

\clearpage
\begin{minipage}{\textwidth}
\includepdf[pages=1,scale=0.95,pagecommand={}]{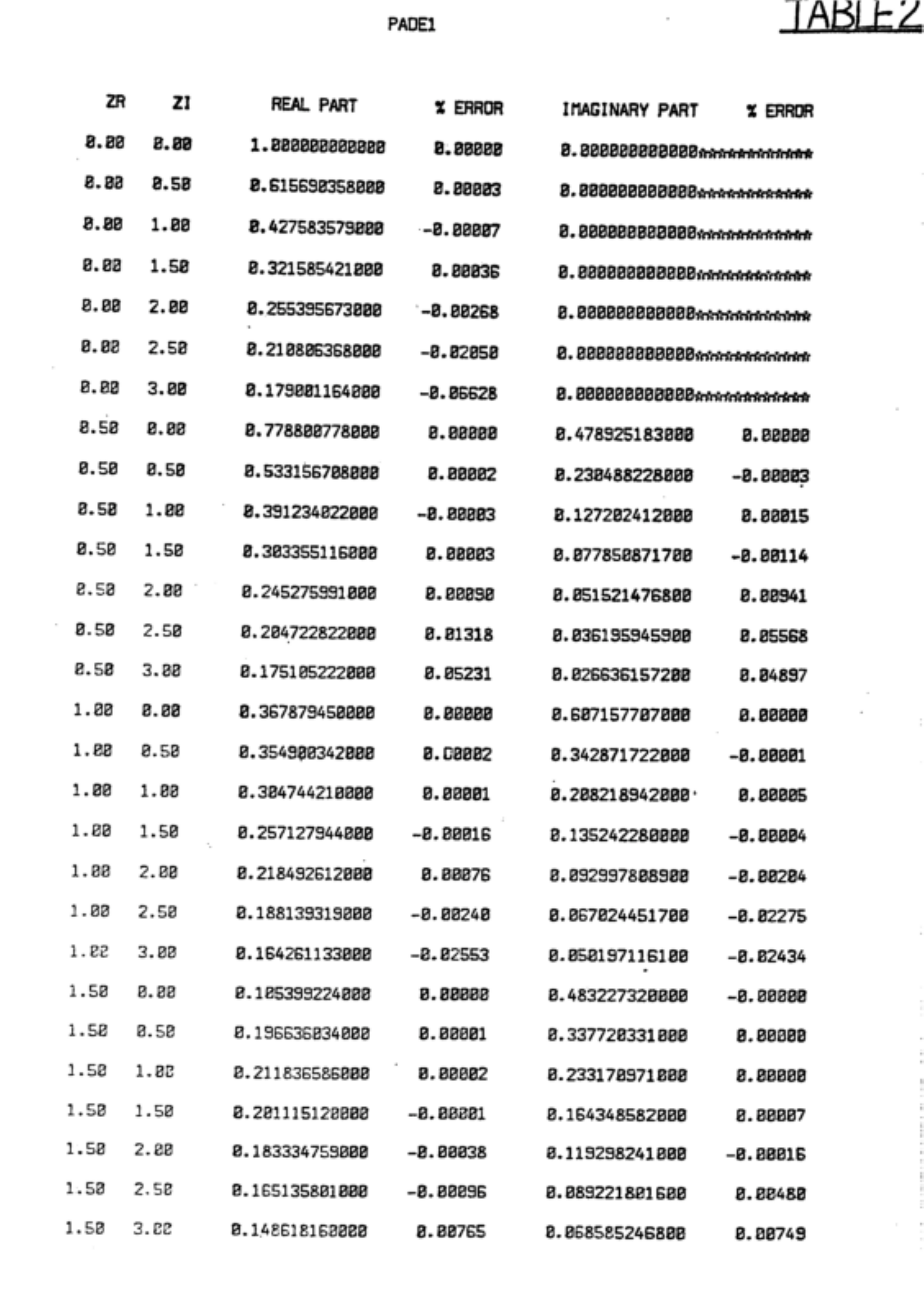} 
\end{minipage}

\clearpage
\begin{minipage}{\textwidth}
\includepdf[pages=1,scale=0.95,pagecommand={}]{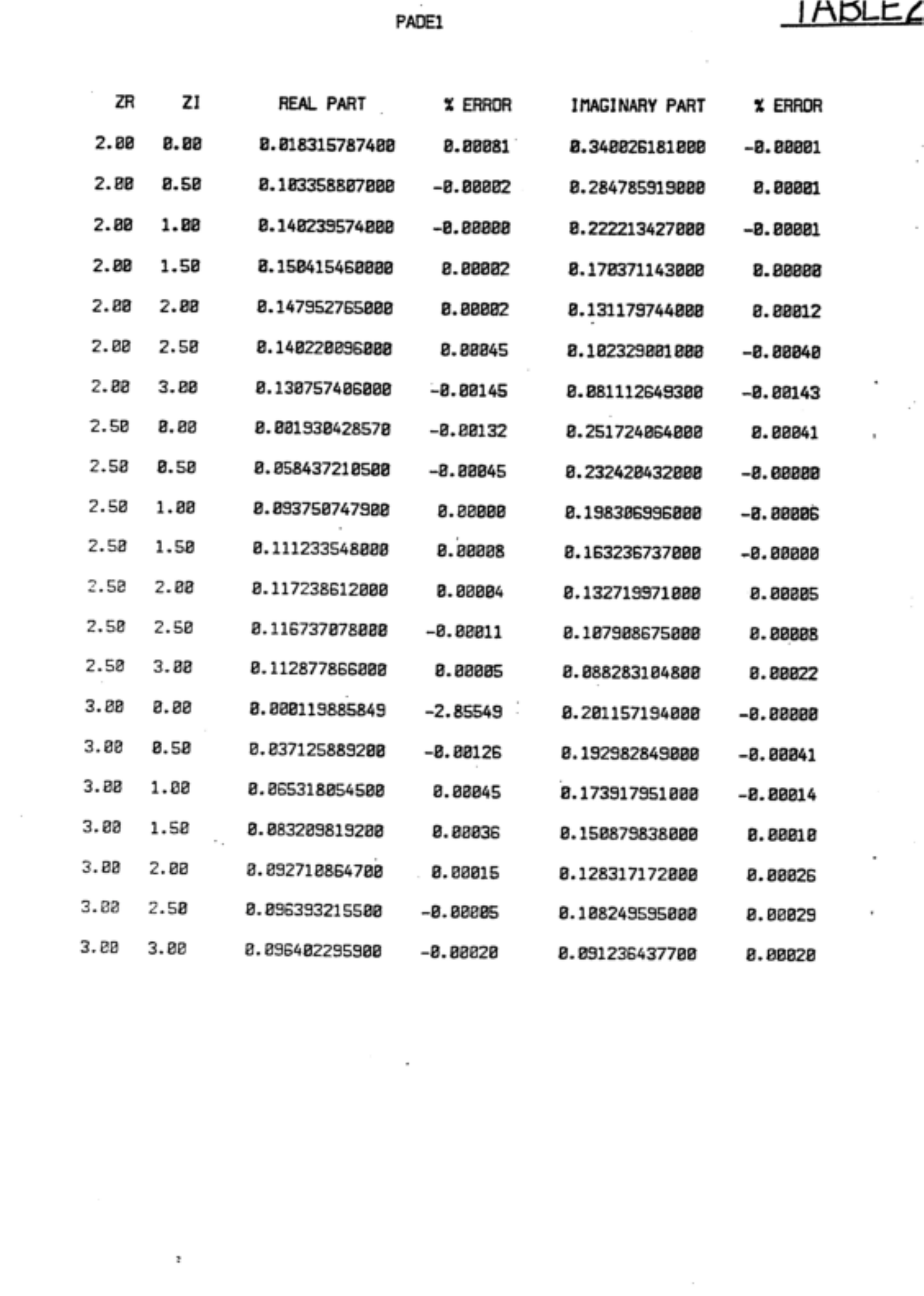} 
\end{minipage}

\clearpage
\begin{minipage}{\textwidth}
\includepdf[pages=1,scale=0.95,pagecommand={}]{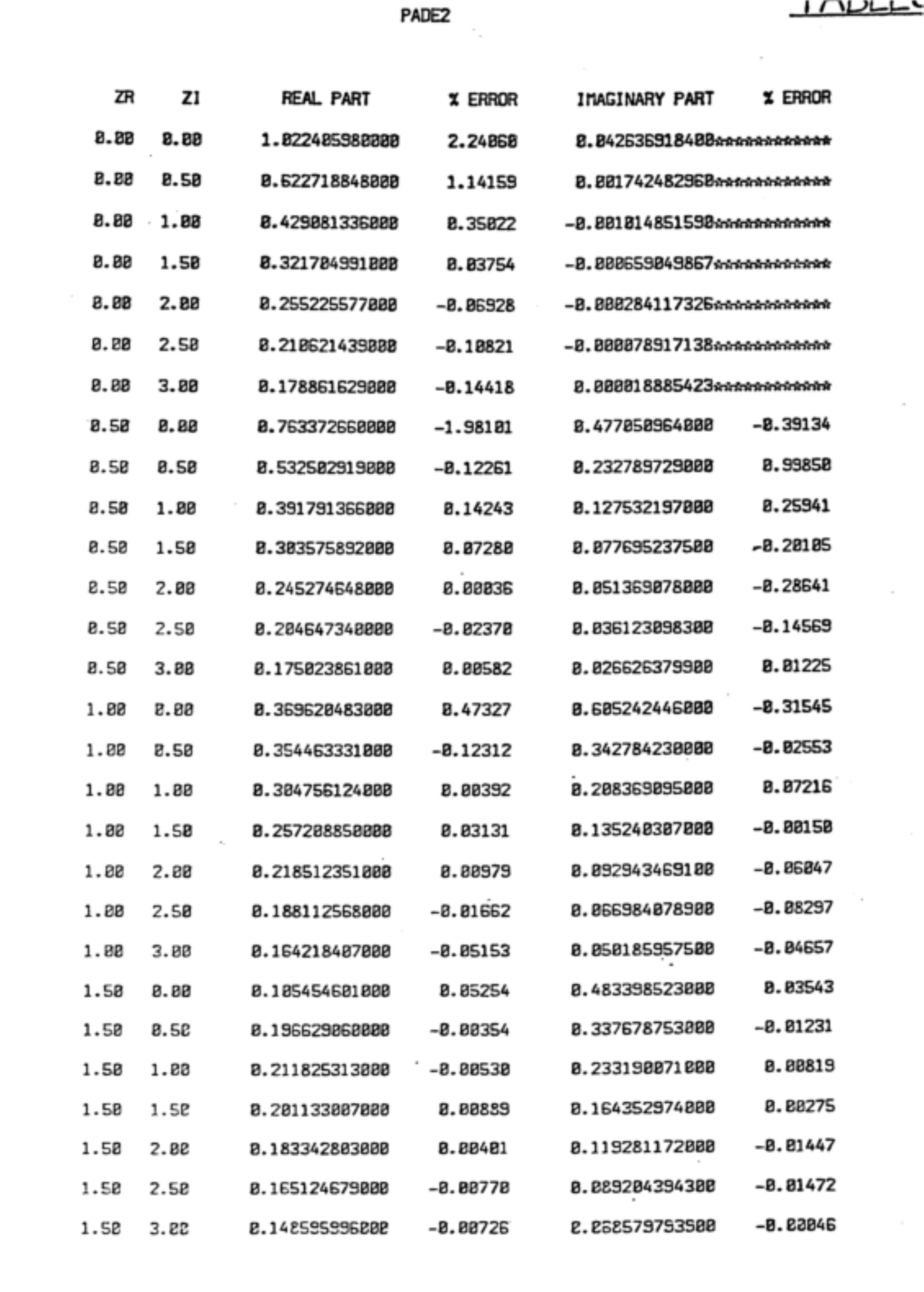} 
\end{minipage}

\clearpage
\begin{minipage}{\textwidth}
\includepdf[pages=1,scale=0.95,pagecommand={}]{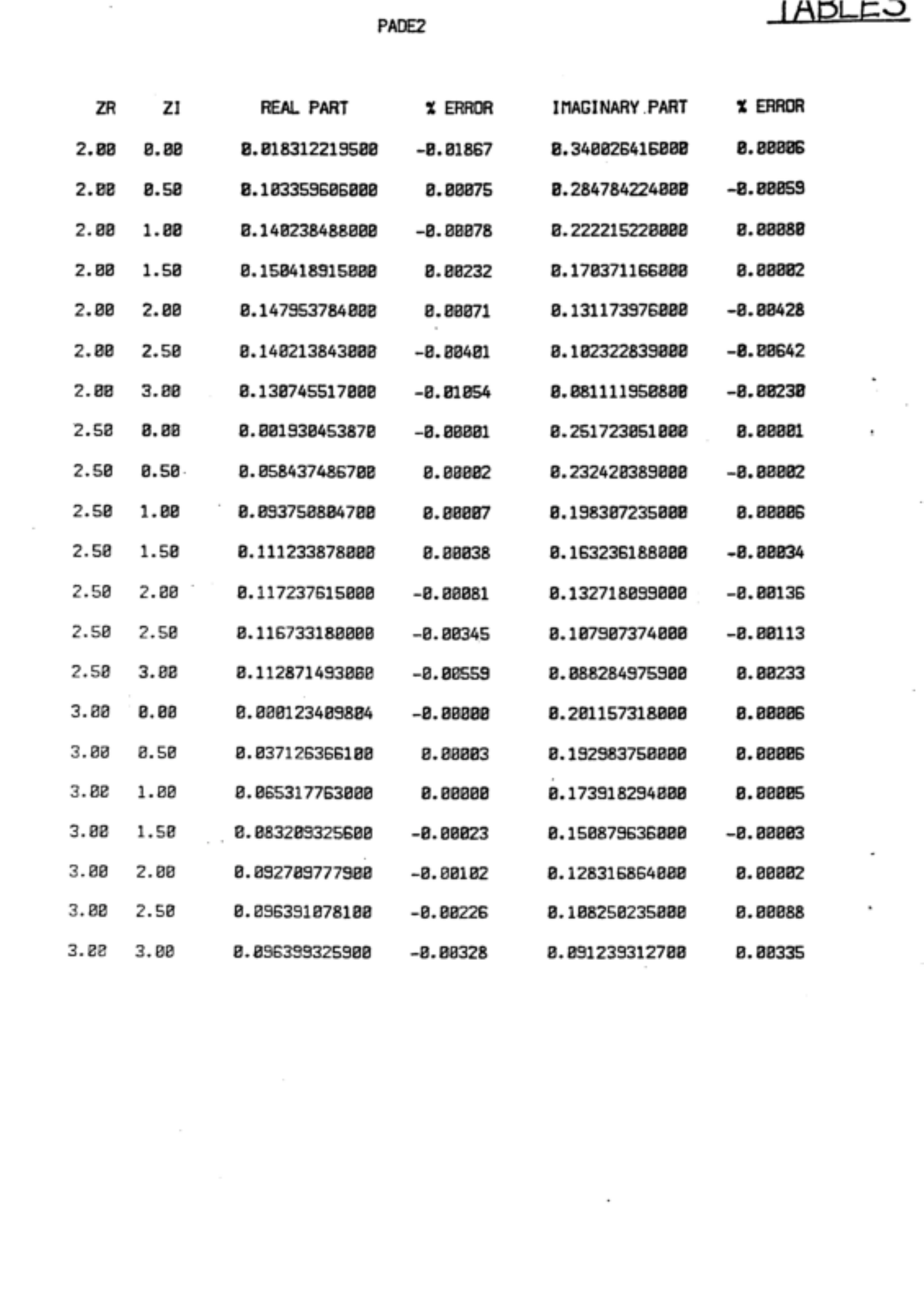} 
\end{minipage}

\clearpage
\begin{minipage}{\textwidth}
\includepdf[pages=1,scale=0.95,pagecommand={}]{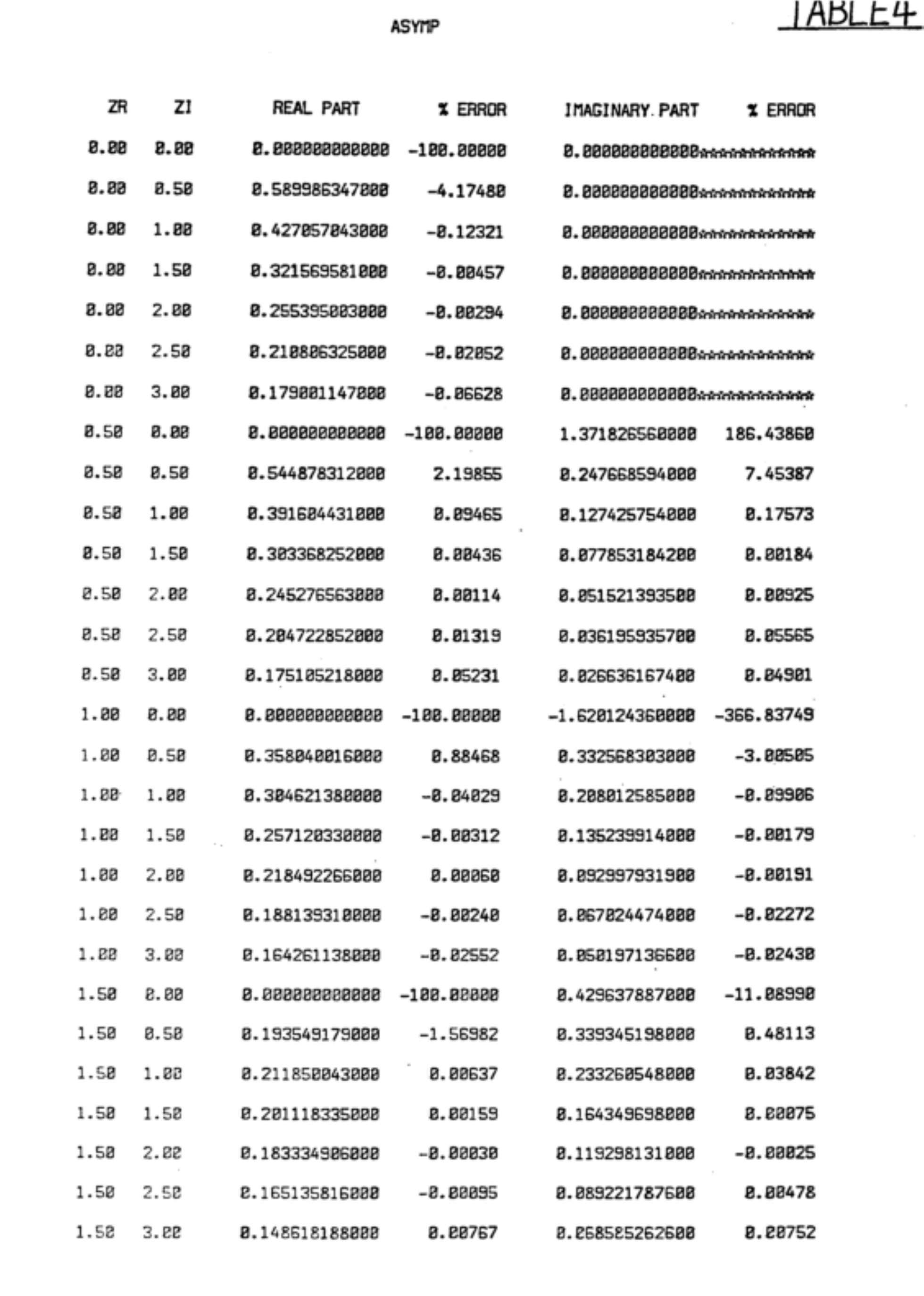} 
\end{minipage}

\clearpage
\begin{minipage}{\textwidth}
\includepdf[pages=1,scale=0.95,pagecommand={}]{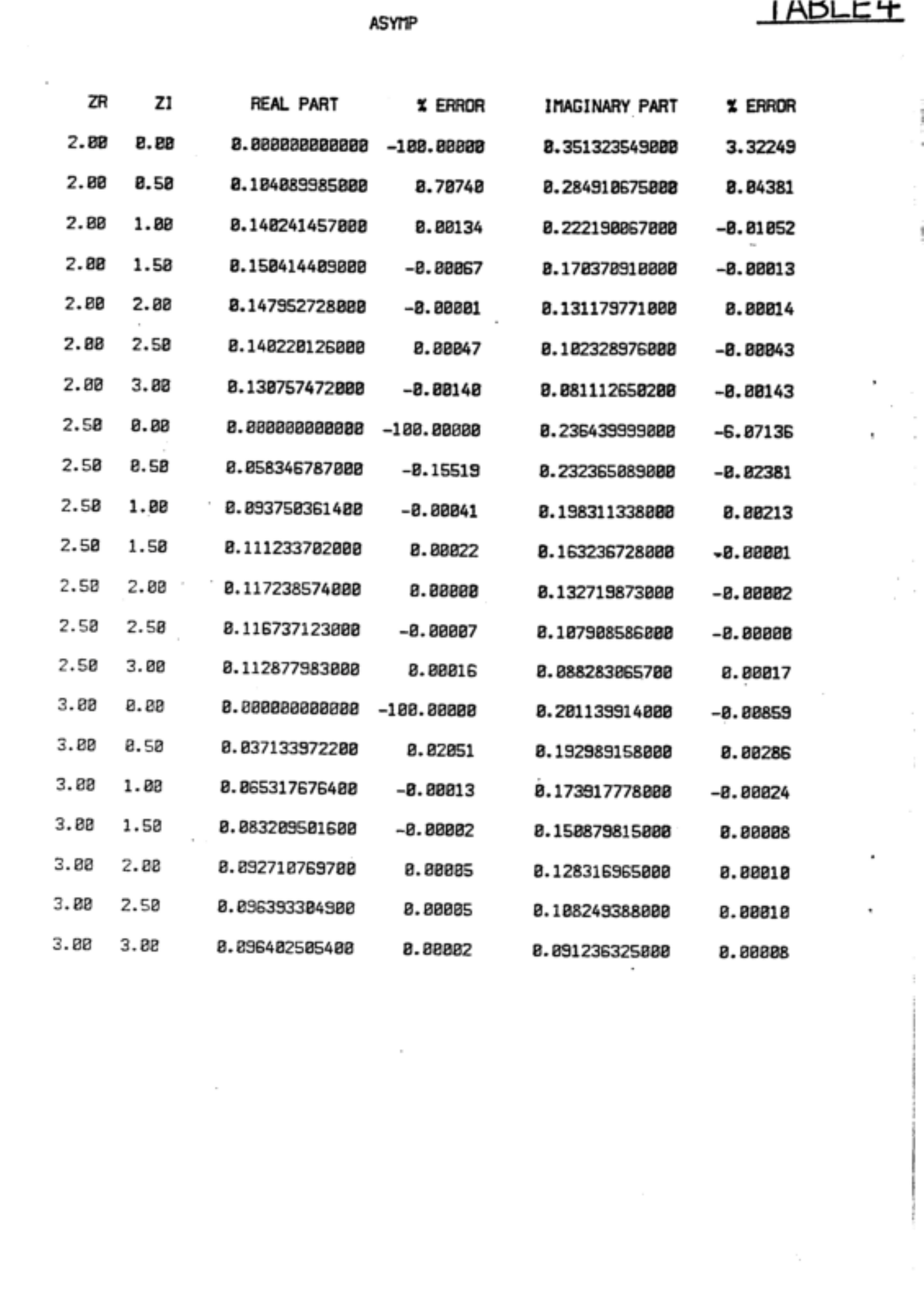} 
\end{minipage}

\clearpage
\begin{minipage}{\textwidth}
\includepdf[pages=1,scale=0.95,pagecommand={}]{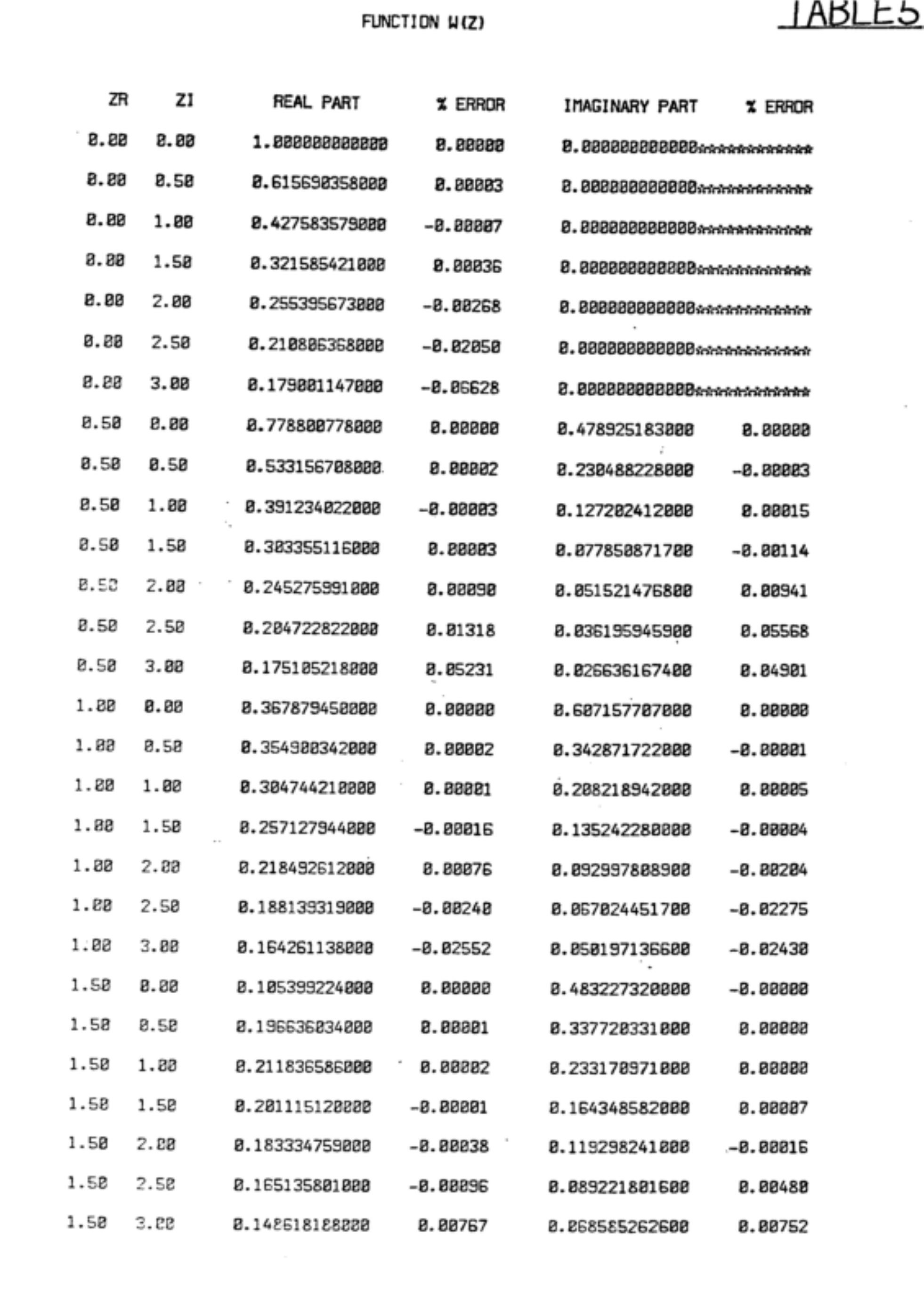} 
\end{minipage}

\clearpage
\begin{minipage}{\textwidth}
\includepdf[pages=1,scale=0.95,pagecommand={}]{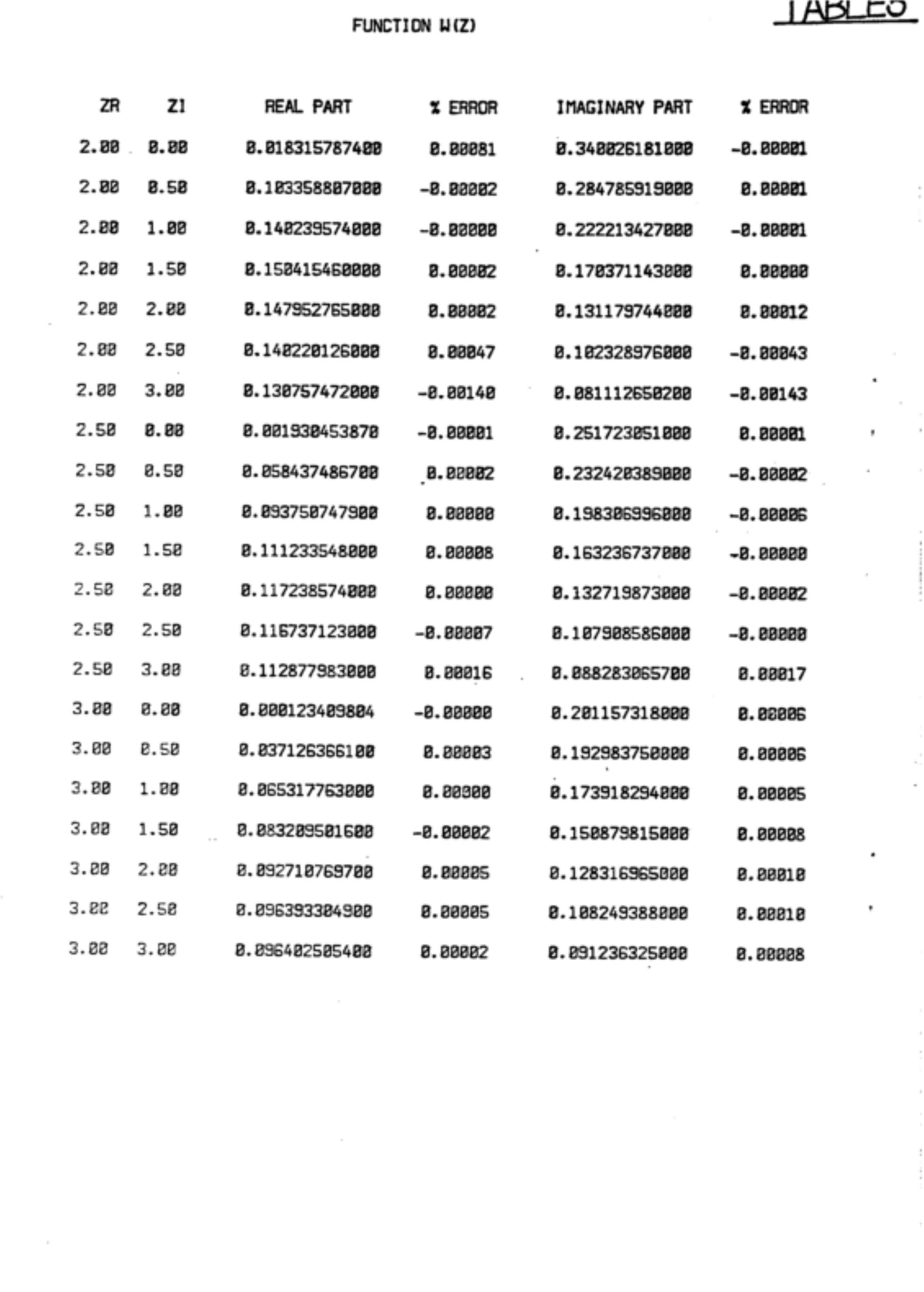} 
\end{minipage}

\clearpage
\begin{minipage}{\textwidth}
\includepdf[pages=1,scale=0.95,pagecommand={}]{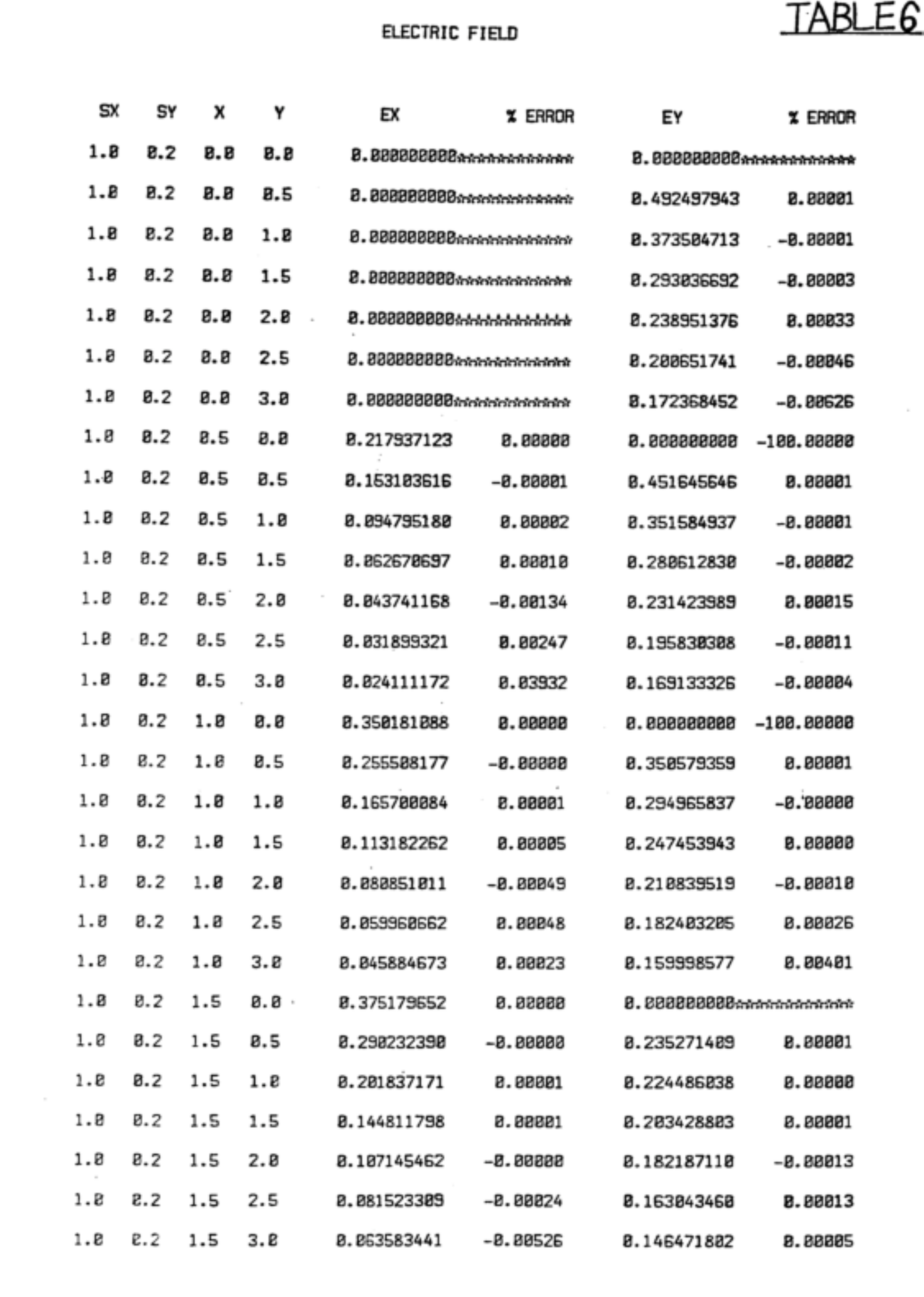} 
\end{minipage}

\clearpage
\begin{minipage}{\textwidth}
\includepdf[pages=1,scale=0.95,pagecommand={}]{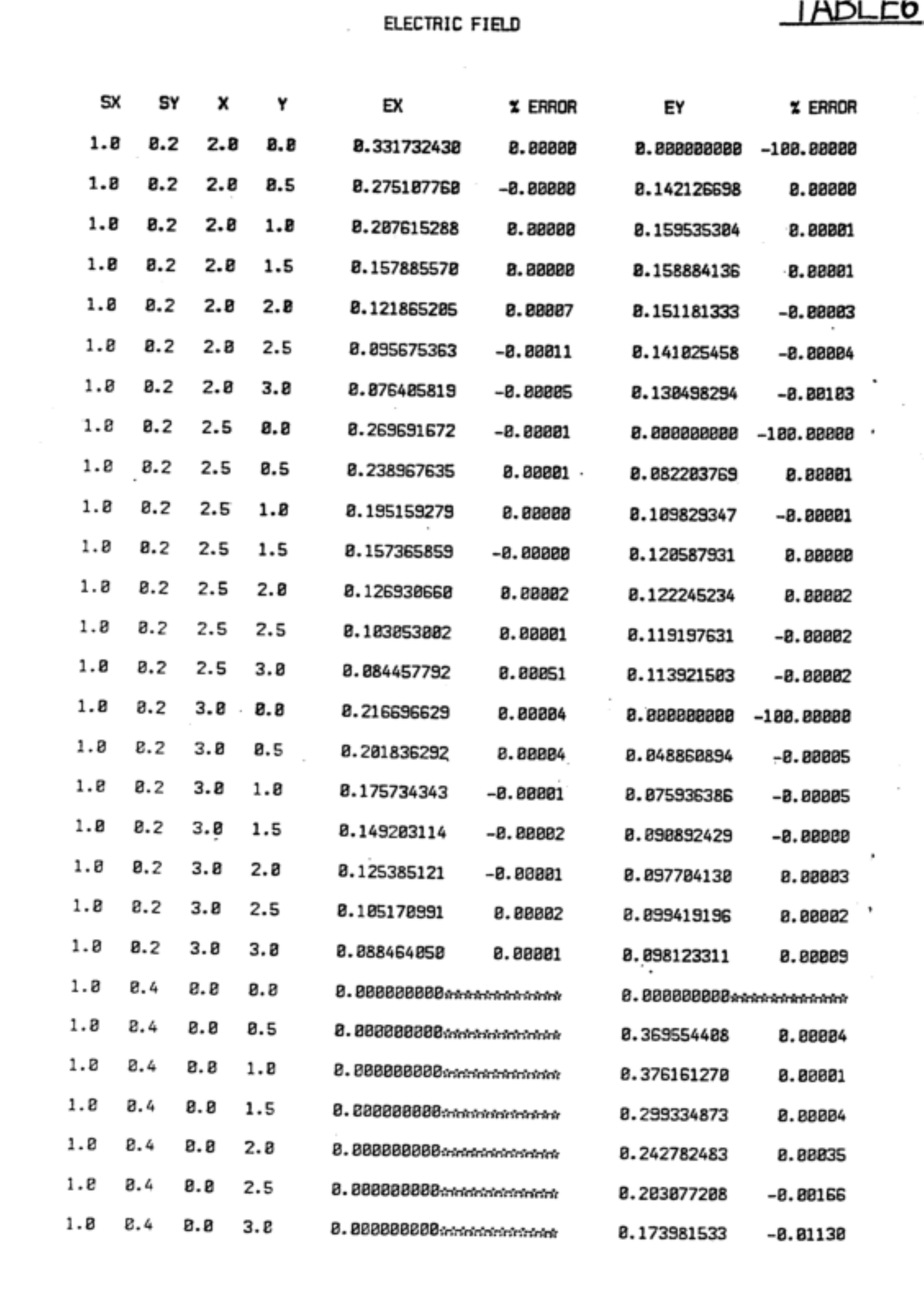} 
\end{minipage}

\clearpage
\begin{minipage}{\textwidth}
\includepdf[pages=1,scale=0.95,pagecommand={}]{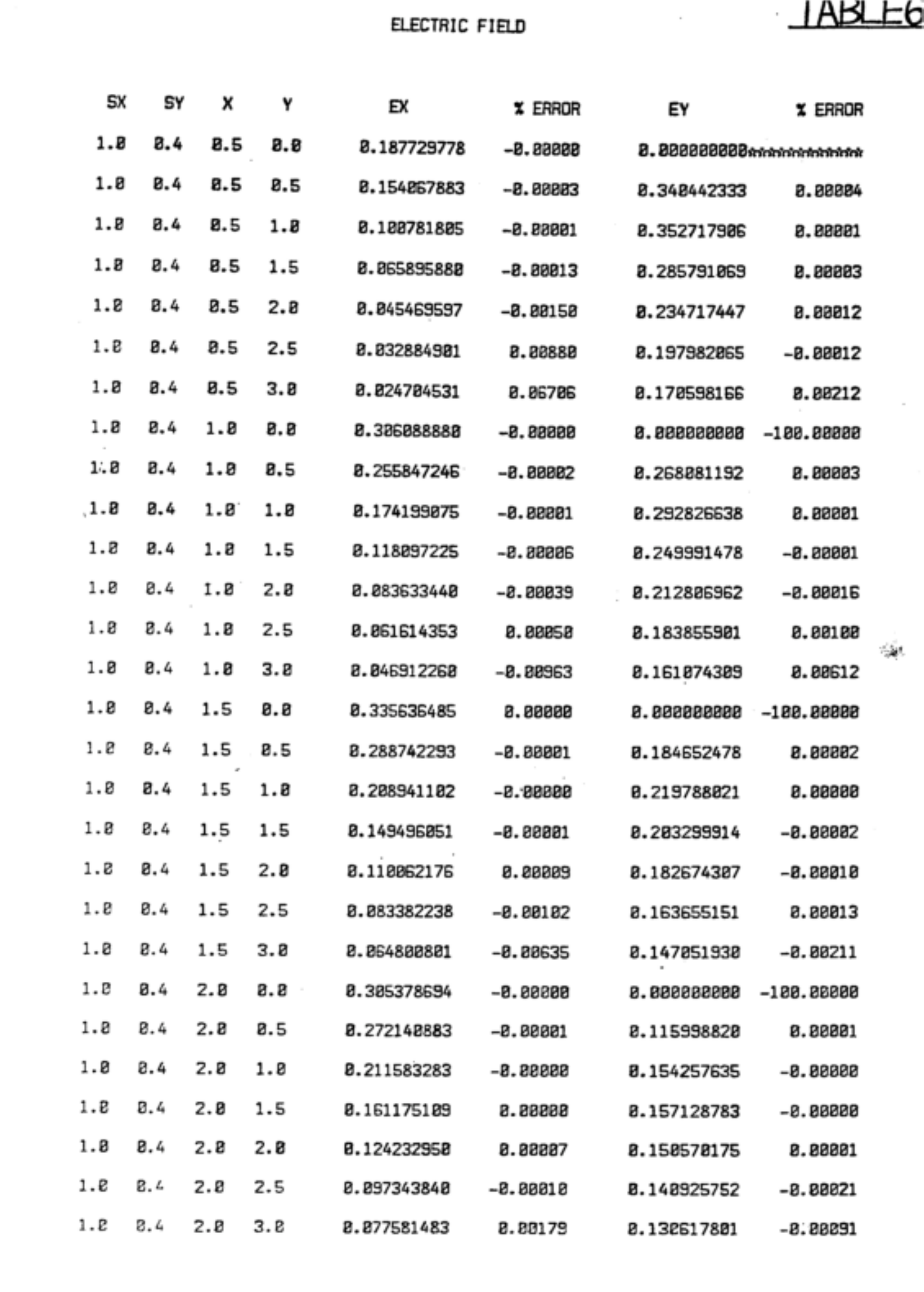} 
\end{minipage}

\clearpage
\begin{minipage}{\textwidth}
\includepdf[pages=1,scale=0.95,pagecommand={}]{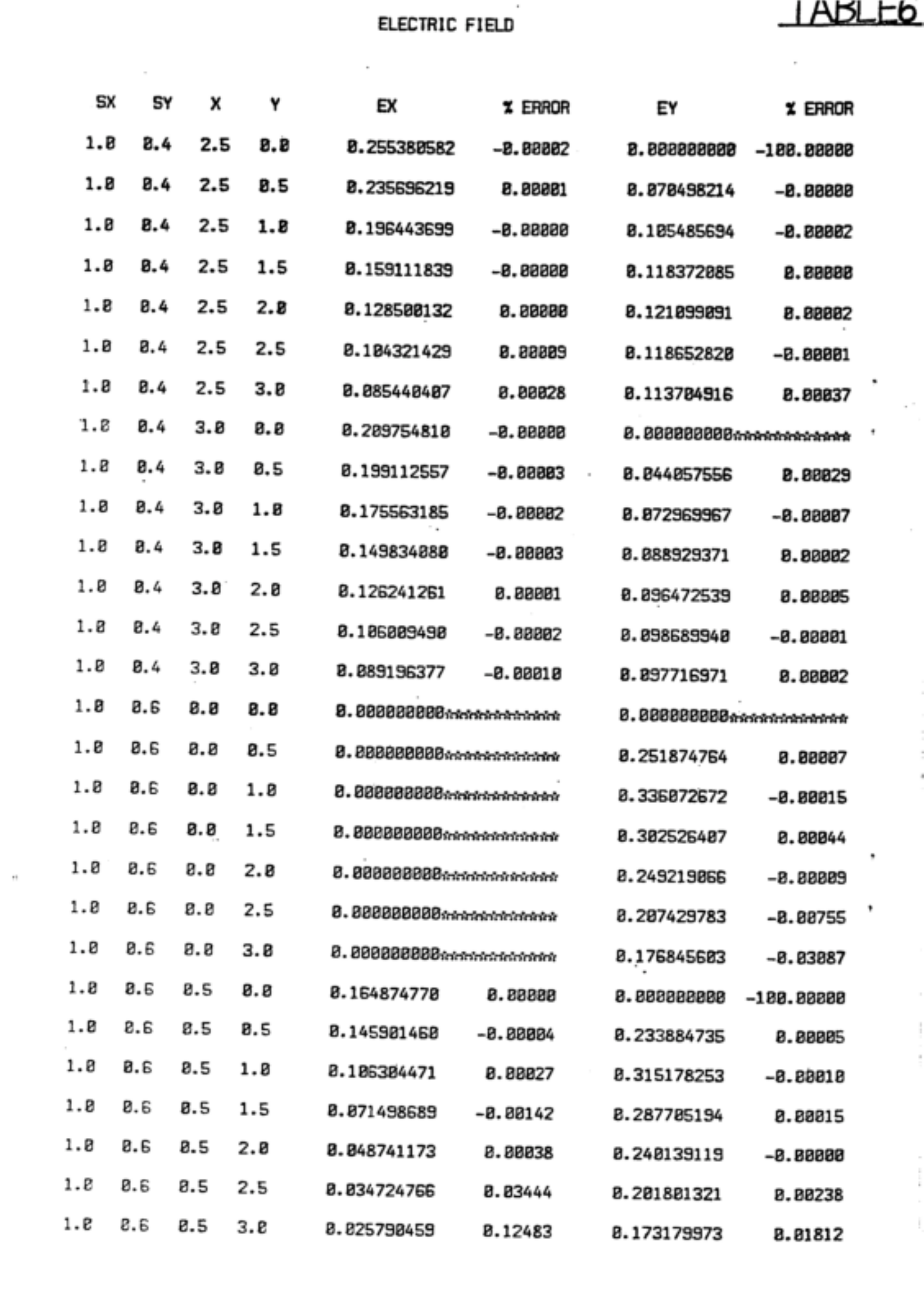} 
\end{minipage}

\clearpage
\begin{minipage}{\textwidth}
\includepdf[pages=1,scale=0.95,pagecommand={}]{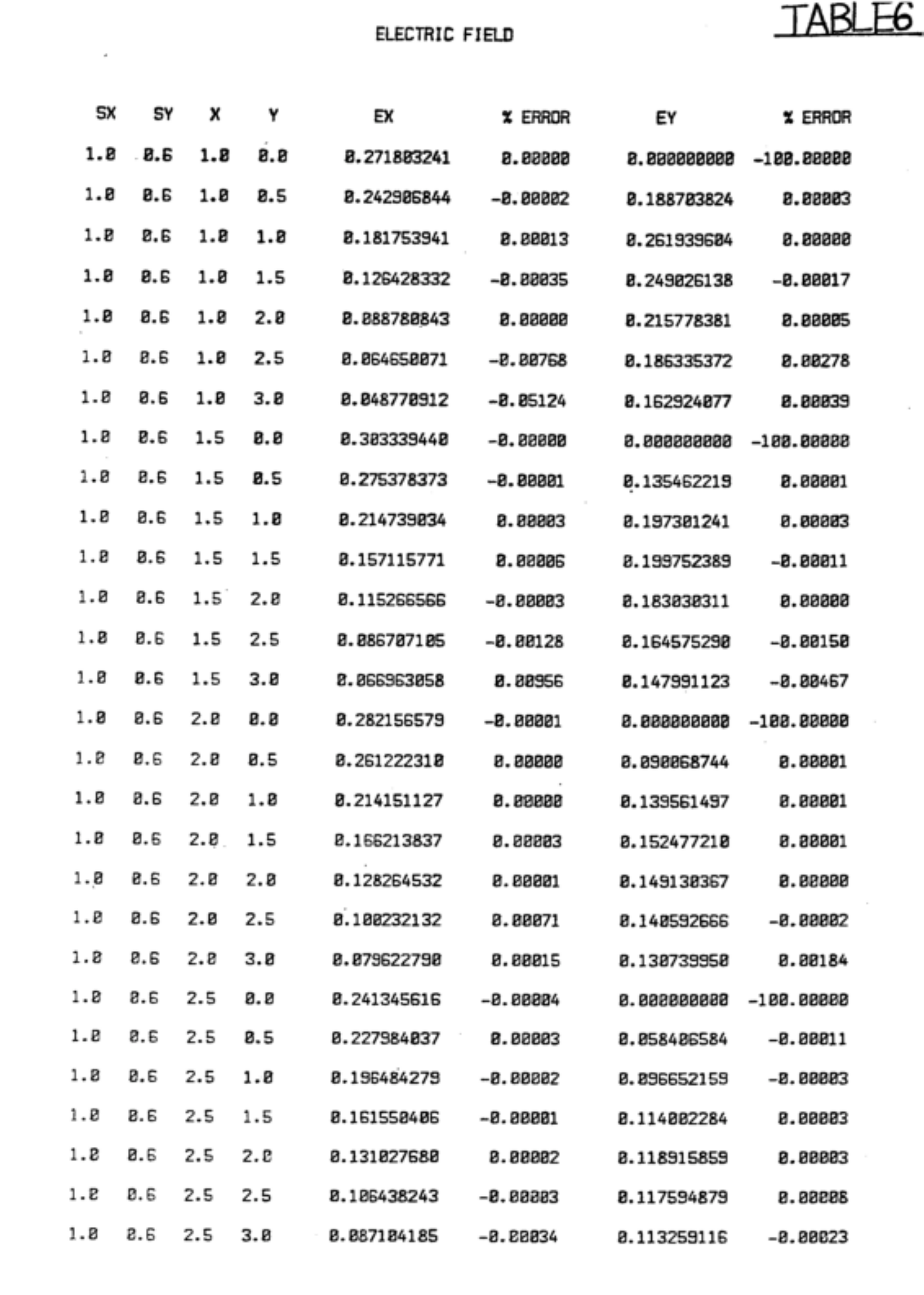} 
\end{minipage}

\clearpage
\begin{minipage}{\textwidth}
\includepdf[pages=1,scale=0.95,pagecommand={}]{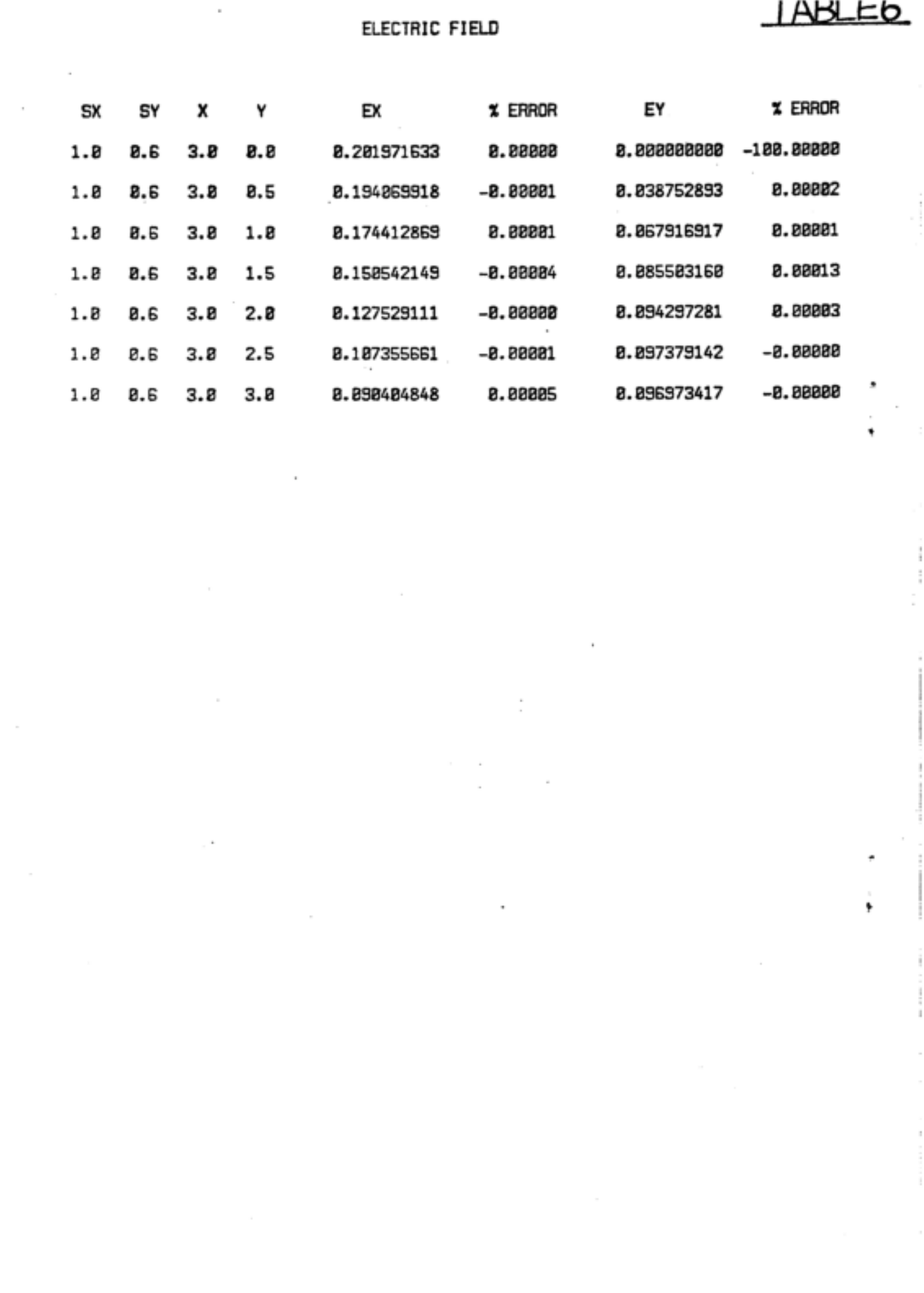} 
\end{minipage}

\clearpage
\section{Recreated Table~1, WEXCT\label{sec:WEXCT}}
 \newcolumntype{P}[1]{>{          \arraybackslash}p{#1}}
 \newcolumntype{Q}[1]{>{\centering\arraybackslash}p{#1}}
\begin{table}[htbp]\footnotesize
 \begin{tabular}{|P{0.5cm} P{1.5cm} P{3.5cm} Q{3.5cm}|}
 \hline
 zr & zi & \centering wr & wi \\ [0.5ex] 
 \hline\hline
0.00 & 0.00 &    1.000000000000000 & 0.000000000000000 \\
0.00 & 0.50 &    0.615690177904398 & 0.000000000000000 \\
0.00 & 1.00 &    0.427583868928526 & 0.000000000000000 \\
0.00 & 1.50 &    0.321584289440764 & 0.000000000000000 \\
0.00 & 2.00 &    0.255402534123747 & 0.000000000000000 \\
0.00 & 2.50 &    0.210849597766694 & 0.000000000000000 \\
0.00 & 3.00 &    0.179119883122594 & 0.000000000000000 \\
 \hline
0.50 & 0.00 &    0.778800783071405 & 0.478925159573737 \\
0.50 & 0.50 &    0.533156591133704 & 0.230488280661727 \\
0.50 & 1.00 &    0.391234138767048 & 0.127202207682148 \\
0.50 & 1.50 &    0.303355049877176 & 0.077851740528118 \\
0.50 & 2.00 &    0.245273758511286 & 0.051516615437352 \\
0.50 & 2.50 &    0.204695834494986 & 0.036175795560759 \\
0.50 & 3.00 &    0.175013660760233 & 0.026623117998033 \\
 \hline
1.00 & 0.00 &    0.367879441171442 & 0.607157688945692 \\
1.00 & 0.50 &    0.354900293344541 & 0.342871755762348 \\
1.00 & 1.00 &    0.304744150133933 & 0.208218830492453 \\
1.00 & 1.50 &    0.257128339312765 & 0.135242331276541 \\
1.00 & 2.00 &    0.218490958441978 & 0.092999718081428 \\
1.00 & 2.50 &    0.188143831076063 & 0.067039707747043 \\
1.00 & 3.00 &    0.164303072898013 & 0.050209338851178 \\
 \hline
1.50 & 0.00 &    0.105399224561864 & 0.483227316693744 \\
1.50 & 0.50 &    0.196636025659636 & 0.337720324630445 \\
1.50 & 1.00 &    0.211836548282862 & 0.233170966384206 \\
1.50 & 1.50 &    0.201115136771154 & 0.164348461581649 \\
1.50 & 2.00 &    0.183335452017590 & 0.119298431859613 \\
1.50 & 2.50 &    0.165137393190194 & 0.089217522723435 \\
1.50 & 3.00 &    0.148606780170188 & 0.068580103542200 \\
 \hline
2.00 & 0.00 &    0.018315638888734 & 0.340026207603973 \\
2.00 & 0.50 &    0.103358822002007 & 0.284785879309291 \\
2.00 & 1.00 &    0.140239573816261 & 0.222213438219392 \\
2.00 & 1.50 &    0.150415415022916 & 0.170371132299577 \\
2.00 & 2.00 &    0.147952737100350 & 0.131179589036453 \\
2.00 & 2.50 &    0.140219456205913 & 0.102329400148157 \\
2.00 & 3.00 &    0.130759301196336 & 0.081113815128269 \\
 \hline
2.50 & 0.00 &    0.001930454136228 & 0.251723017570233 \\
2.50 & 0.50 &    0.058437471245225 & 0.232420429753165 \\
2.50 & 1.00 &    0.093750740957915 & 0.198307111906521 \\
2.50 & 1.50 &    0.111233455711503 & 0.163236744644939 \\
2.50 & 2.00 &    0.117238559619234 & 0.132719893761845 \\
2.50 & 2.50 &    0.116737205072723 & 0.107908588497039 \\
2.50 & 3.00 &    0.112877805272293 & 0.088282913440885 \\
\hline
3.00 & 0.00 &    0.000123409804087 & 0.201157191681505 \\
3.00 & 0.50 &    0.037126355895963 & 0.192983630876662 \\
3.00 & 1.00 &    0.065317757536137 & 0.173918193589552 \\
3.00 & 1.50 &    0.083209506894530 & 0.150879672266300 \\
3.00 & 2.00 &    0.092710730944052 & 0.128316849682324 \\
3.00 & 2.50 &    0.096393255412096 & 0.108249277644066 \\
3.00 & 3.00 &    0.096402467057972 & 0.091236235688546 \\
 \hline
\end{tabular}
\centering\caption {WEXCT (re-calculated)}
\end{table}

\end{document}